\documentclass[12pt]{iopart}

\usepackage{iopams}
\usepackage{epsf}

\newcommand{\bd}{\begin{displaymath}}
\newcommand{\ed}{\end{displaymath}}
\newcommand{\be}{\begin{equation}}
\newcommand{\ee}{\end{equation}}
\newcommand{\ba}{\begin{eqnarray}}
\newcommand{\ea}{\end{eqnarray}}

\begin{document}

\topical[Quasielastic He atom scattering from surfaces]
{Quasielastic He atom scattering from surfaces:\\
A stochastic description of the dynamics of interacting adsorbates}

\author{R Mart\'{\i}nez-Casado$^{\dag,*}$, J L Vega$^{\ddag,*}$,
A S Sanz$^*$, and\\ S Miret-Art\'es$^*$}

\address{$^\dag$Lehrstuhl f\"ur Physikalische Chemie I,
Ruhr-Universit\"at Bochum, D-44801 Bochum, Germany.}

\address{$^\ddag$Biosystems Group, School of Computing,
University of Leeds, Leeds LS2 9JT, United Kingdom}

\address{$^*$Instituto de Matem\'aticas y F\'{\i}sica Fundamental,
Consejo Superior de Investigaciones Cient\'{\i}ficas,
Serrano 123, 28006 Madrid, Spain.}

\eads{\mailto{ruth@imaff.cfmac.csic.es},
\mailto{jlvega@imaff.cfmac.csic.es},
\mailto{asanz@imaff.cfmac.csic.es},
\mailto{s.miret@imaff.cfmac.csic.es}}

\begin{abstract}
The study of diffusion and low frequency vibrational motions of
particles on metal surfaces is of paramount importance; it provides
valuable information on the nature of the adsorbate-substrate and the
substrate-substrate interactions.
In particular, the experimental broadening observed in the diffusive
peak with increasing coverage is usually interpreted in terms of
a dipole-dipole like interaction among adsorbates via extensive
molecular dynamics calculations within the Langevin framework.
Here we present an alternative way to interpret this broadening by
means of a purely stochastic description, namely the interacting
single adsorbate approximation, where two noise sources are considered:
(1) a Gaussian white noise accounting for the surface friction and
temperature, and (2) a white shot noise replacing the interaction
potential between adsorbates. Standard Langevin numerical simulations
for flat and corrugated surfaces (with a separable potential)
illustrate the dynamics of Na
atoms on a Cu(100) surface which fit fairly well to the
analytical expressions issued from simple models (free particle and
anharmonic oscillator) when the Gaussian approximation is assumed.
A similar broadening is also expected for the frustrated translational
mode peaks.
\end{abstract}

\submitto{\JPCM}

\pacs{03.65.-w, 03.65.Ta, 79.20.Rf}

\tableofcontents

\maketitle


\section{Introduction}
\label{sec1}

Diffusion and low-frequency vibrational motion of atomic and
molecular adsorbates on surfaces are two very important and elementary
dynamical processes.
They provide valuable information on the nature of the
adsorbate-substrate and adsorbate-adsorbate interactions.
Moreover, a good understanding of these processes also constitutes a
preliminary step in the study of more complicated phenomena in surface
science, such as heterogeneous catalysis, crystal growth, lubrication,
chemical vapor deposition, associative desorption, etc.
Among the different experimental techniques used to study
these processes and extract interaction potentials,
\cite{Gomer,Ehrlich,toennies4,salva,jardine1},
the {\it quasielastic He atom scattering} (QHAS)
\cite{toennies4,toennies1,toennies2,toennies3} is considered as the
surface science analogue of quasielastic neutron scattering, which has
been widely and successfully applied to analyze diffusion in the bulk
\cite{vanhove,vineyard}.

Diffusion times and thus diffusion coefficients may vary by orders of
magnitude, depending on the temperature of the surface and the nature
of the adsorbate-surface interaction (for a recent review, see for
example, reference~\cite{salva}).
Scanning tunnelling microscopy (STM) and field ion microscopy (FIM) are
especially useful for slow diffusion, where the time between jumps of
the adatom is of the order of seconds.
These measurements provide a direct observation of the
diffusion process. QHAS is being used to determine diffusion processes
in which the time between jumps is of the order of microseconds.
In contrast with STM and FIM, QHAS provides information on the
diffusion dynamics only indirectly, but it is one of the techniques
where reliable adsorbate-adsorbate and adsorbate-substrate
interactions can be obtained since the underlying theory  is much more
intrincate. Most of theoretical treatments applied to interpret QHAS
results assume that the interaction of He atoms with the adsorbates
and the surface is negligible. As far as we know, a general study on
how the probe particles can distort the diffusion process is not
available yet.

In the QHAS technique, time-of-flight measurements are converted
to energy transfer spectra from which a wide energy range
can be spanned and several peaks are observed.
The prominent peak around the zero energy transfer, namely the
{\it quasielastic peak} (Q peak), gives information about
the adsorbate diffusion process.
Additional weaker peaks at low energy transfers around the Q peak are
also observed.
These peaks are attributed to the parallel (to the surface) frustrated
translational motion of the adsorbates (T modes) and also to the
excitations of surface phonons (at positive energy transfers we
have creation processes and annihilation ones at negative energy
transfers).
The measurable quantity experimentally is the so-called {\it dynamic
structure factor} or {\it scattering law}, which gives the line shapes
of all those elementary processes, in particular those corresponding
to the Q and T peaks.
The dynamic structure factor provides information about the dynamics
and structure of the adsorbates through particle distribution
functions, the latter being related to the nature of the
adsorbate-substrate and the adsorbate-adsorbate interactions.

At low coverages the interactions among adsorbates can be ignored,
this allowing to work within the so-called {\it single adsorbate
approximation}.
In this case, diffusion (or {\it self-diffusion}) is characterized
by only taking into account the adsorbate-substrate interaction
which  is usually described by a stochastic model (Brownian-like
motion) with two contributions:
(1) the deterministic, phenomenological adiabatic potential, $V$, which
models the interaction at $T=0$; and (2) a stochastic force, $R_G(t)$,
(Gaussian white noise) accounting for the vibrational effects induced
by the temperature on the surface lattice atoms (and therefore on the
adsorbates). In this framework, the dynamics is then carried out by
solving the standard Langevin equation
\cite{salva,JLvega0,JLvega1,eli,sancho}.
In this type of dynamical calculation the parameters defining the
interaction potential are chosen very often to obtain good agreement
with the experimental data.
This way of proceeding has the disadvantage that such potential
functions are not unique, since there can be a multiplicity of forms
leading to the same results; only {\it ab initio} calculations could
provide, in principle, unique potential functions.

When dealing with higher coverages, adsorbate-adsorbate interactions
can no longer be neglected.
In this case, the adsorbate-surface interaction can still be described
as before, but pairwise potential functions accounting for the
adsorbate-adsorbate interactions are usually introduced into
Langevin molecular dynamics (LMD) simulations \cite{toennies3}.
An alternative approach is to consider a purely stochastic description
for these interactions, as we have shown elsewhere \cite{prl-I,pre-I}.
This description, which we have denominated the {\it interacting
single adsorbate approximation}, is a (fully) Langevin approach based
on the theory of spectral-line collisional broadening developed by
Van Vleck and Weisskopf \cite{vvleck-weiss} and the elementary kinetic
theory of gases \cite{mcquarrie}.
This new approach explains fairly well experimental results where
broadening is observed with increasing coverage in both the Q and
the T-mode peaks \cite{toennies3}.
The motion of a single adsorbate is modelled by a
series of random pulses within a Markovian regime (i.e., pulses of
relatively short duration in comparison with the system relaxation);
the pulses simulate the collisions with other adsorbates.
In particular, we describe the adsorbate-adsorbate collisions by
means of a {\it white shot noise} as a limiting case of a colored shot
noise \cite{gardiner}.
The concept of shot noise has been applied to study thermal ratchets
\cite{hanggi1}, mean first passage times \cite{iturbe}, or jump
probabilities \cite{tommei}; an interesting, recent account on general
colored noise in dynamical systems can be found in \cite{hanggi2}.

In this work we present a detailed analytical and numerical description
of the effects arising from both Gaussian white noise and shot noise
when applied to surface diffusion problems.
In particular, we show how the kinds of process taking place on a
surface can be properly described analytically by means of a very
simple theoretical framework, where the interplay between diffusion
and low vibrational motions results very apparent.
Accordingly, we have organized this work as follows.
The description of the elements involved in the theoretical analysis
of line shapes as well as the mathematics related to the analytical
results are given in section~\ref{sec2}.
In section~\ref{sec3} we provide a numerical illustration of the
surface dynamics for different types of models to give a broad
perspective of our approach.
Finally, the main conclusions arising from this work as well as some
future trends are given in section~\ref{sec4}.


\section{Stochastic approaches to surface diffusion}
\label{sec2}


\subsection{Dynamic structure factor and intermediate scattering
function}
\label{sec2.1}

In QHAS experiments one is usually interested in the {\it differential
reflection coefficient}, which can be expressed as
\be
 \frac{d^2 \mathcal{R} (\Delta {\bf K}, \omega)}{d\Omega d\omega}
  = n_d \mathcal{F} S(\Delta {\bf K}, \omega)
 \label{eq:DRP}
\ee
in analogy to scattering of slow neutrons by crystals and
liquids \cite{vanhove,vineyard}.
This magnitude gives the probability that the probe He atoms scattered
from the diffusing collective reach a certain solid angle $\Omega$ with
an energy exchange $\hbar\omega =E_f - E_i$ and a parallel (to the
surface) momentum transfer $\Delta {\bf K} = {\bf K}_f - {\bf K}_i$.
In equation~(\ref{eq:DRP}), $n_d$ is the (diffusing) surface
concentration of adparticles; $\mathcal{F}$ is the {\it atomic form
factor}, which depends on the interaction potential between the probe
atoms in the beam and the adparticles on the surface; and
$S(\Delta {\bf K},\omega)$ is the {\it dynamic structure factor} or
{\it scattering law}, which gives the line shapes of the Q and the
T-mode peaks ---other peaks can also be
present, such as the inelastic ones related to surface phonon
excitations--- and provides a complete information about the dynamics
and structure of the adsorbates through particle distribution
functions \cite{vanhove,vineyard}.
Experimental information about long distance correlations is obtained
from the scattering law when considering small values of $\Delta {\bf
 K}$, while information on long time correlations is available at
small energy transfers $\hbar \omega$.

Surface diffusion is a dynamical problem that can be tackled by means
of classical mechanics for heavy adparticles.
Thus, let us consider an ensemble of classical interacting particles
on a surface.
Their distribution functions are described by means of the so-called
van Hove or time-dependent pair correlation function $G({\bf R},t)$
\cite{vanhove}.
This function is related to the dynamic structure factor as
\be
 S(\Delta {\bf K}, \omega)
  = \int \! \! \! \int G({\bf R},t)
  e^{i(\Delta {\bf K} \cdot {\bf R} -\omega t)}
   \ \! d{\bf R} \ \! dt .
 \label{eq:DRP2}
\ee
Given an adparticle at the origin at some arbitrary initial time,
$G({\bf R},t)$ represents the average probability for finding a
particle (the same or another one) at the surface position
${\bf R} = (x,y)$ at a time $t$.
Note that this function is a generalization of the well known pair
distribution function $g({\bf R})$ from statistical mechanics
\cite{mcquarrie,Hansen}, since it provides information about the
interacting particle dynamics.

Depending on whether correlations of an adparticle with itself or
with another one are considered, a distinction can be made between
the {\it self} correlation function, $G_s({\bf R},t)$, and the
{\it distinct} correlation function, $G_d({\bf R},t)$.
The full pair correlation function can then be expressed as
\begin{equation}
 G({\bf R},t) = G_s({\bf R},t) + G_d({\bf R},t) .
 \label{gtotal}
\end{equation}
According to its definition, $G_s({\bf R},t)$ is peaked at $t = 0$
and approaches zero as time increases because the adparticle
{\it loses} correlation with itself. On the other hand,
$G_d({\bf R},t) \equiv g({\bf R})$ gives the {\it static pair
correlation function} ---the standard pair distribution function---
at $t=0$, and approaches the mean surface number density $\sigma$ of
diffusing particles as $t \to \infty$.
Accordingly, equation~(\ref{gtotal}) can be split up as
\begin{equation}
 G({\bf R},0) = \delta ({\bf R}) + g({\bf R})
 \label{gtotal0}
\end{equation}
at $t = 0$, and expressed as
\begin{equation}
 G({\bf R},t) \approx \sigma
 \label{gtotalinfty}
\end{equation}
for a homogeneous system with $\|{\bf R}\| \to \infty$ and/or
$t \to \infty$.
At low adparticle concentrations ($\theta \ll 1$), i.e., when
interactions among adsorbates can be neglected because they are far
apart from each other, the main contribution to equation~(\ref{gtotal})
is $G_s$ (particle-particle correlations are negligible and
$G_d \approx 0$).
On the contrary, for high coverages it is expected that $G_d$ presents
a significant contribution to equation~(\ref{gtotal}).

Within this theoretical framework, the dynamic structure factor is
better written as \cite{vanhove}
\be
 S(\Delta {\bf K},\omega) =
  \int e^{-i\omega t} \ \! I(\Delta{\bf K},t) \ \! dt ,
 \label{eq:DSF}
\ee
where
\be
 I(\Delta {\bf K},t) \equiv \frac{1}{N}
  \langle \sum_{j,j'}^N e^{-i\Delta {\bf K} \cdot
   {\bf R}_j (0)} e^{i \Delta {\bf K} \cdot {\bf R}_{j'}(t)} \rangle
 \label{eq:IntSFf}
\ee
is the {\it intermediate scattering function} ---note that this
function is the {\it space} Fourier transform of $G({\bf R},t)$.
In equation~(\ref{eq:IntSFf}) the brackets denote an ensemble average
and ${\bf R}_j (t)$ the trajectory of the $j$th adparticle on the
surface.
This function can again be split up into two sums, distinct ($I_d$) and self
($I_s$), depending on whether the crossing terms are taken into account
or not, respectively.
Following the language used in neutron scattering theory, the
corresponding Fourier transforms of $I$ and $I_s$ give what is
called the coherent scattering law $S(\Delta {\bf K},\omega)$ and
incoherent scattering law $S_s (\Delta {\bf K},\omega)$, respectively.
In QHAS experiments, and with interacting adsorbates, coherent
scattering is always obtained.
The corresponding theoretical interpretation of that scattering is
usually carried out in terms of the Vineyard convolution approximation
\cite{vineyard}, where the distinct correlation function $G_d$ is
expressed as a convolution of the self correlation function $G_s$.
This approximation is known to fail at small distances where the
surface lattice becomes important.
Whereas in  neutron scattering many attempts to improve the convolution
approximation have been developed, within the QHAS context very little
effort has been devoted to this goal.

At finite coverages, one usually distinguishes between two diffusion
coefficients: the tracer diffusion constant ($D_t$) and the collective
diffusion constant ($D_c$) \cite{Gomer}.
The first one refers to the self-diffusion process and focuses on the
motion of a single adsorbate.
On the contrary, $D_c$ is related to the collective motion of all
adsorbates which is governed by Fick's law.
In any case, a Kubo-Green formula relates $D_t$ or $D_c$ with the
velocity autocorrelation function of a
single adsorbate or with the corresponding for the velocity of the
center of mass, respectively.


\subsubsection{The dipole-dipole interaction}
\label{sec2.1.1}

For interacting adsorbates, LMD simulations are widely carried out.
Typically, one solves numerically a system of $N$ coupled Langevin
equations where $N$ denotes the number of adsorbates.
In most of systems, the Markovian-Langevin approximation is assumed
because the Debye energy of the substrate excitations is greater than
the adsorbate T-mode energy and therefore the damping can be
considered as instantaneous (memory effects are negligible).
The adsorbate-adsorbate interaction is also assumed pairwise and is
given by a repulsive dipole-dipole potential.
The so-called Topping's depolarization formula relates the dipole
strength with the coverage \cite{topping}.
This type of interaction is attributed to the electrostatic repulsion
between the dipoles due to the charge transfer from the adatoms to the
substrate.
Numerical discrepancies between the experimental and molecular dynamics
results are found for the broadening of the Q and the T-mode peaks as a
function of the parallel wave vector momentum transfer.
The discussion about the origin of such discrepancies is still an open
problem \cite{toennies3}.


\subsection{The interacting single adsorbate approximation}
\label{sec2.2}

When the coverage increases, the introduction of pairwise potential
functions results in a realistic description of the adsorbate dynamics.
However, there is no a simple manner to handle the resulting
calculations by means of a simple theoretical model, and one can only
proceed by using some suggested fitting functions.
Moreover, carrying out LMD simulations always result in a relatively high
computational cost due to the time spent by the codes in the
evaluation of the forces among particles.
This problem gets worse when working with long-range interactions,
since {\it a priori} they imply that one should consider a relatively
large number of particles in order to get a good numerical simulation.
One way to avoid such inconveniences (interpretative and computational)
could consist in using a simple, realistic stochastic model.
For a good simulation of a diffusion process, one
has to consider very long times in comparison to the timescales
associated with the friction caused by the surface or to the typical
vibrational frequencies observed when the adsorbates keep moving
inside a surface well. This means that there will be a considerably
large number of collisions during the time elapsed in the propagation,
and therefore that, at some point, the past history of the adsorbate
could be irrelevant regarding the properties we are interested in.
This memory loss is a signature of a Markovian dynamical regime, where
adsorbates have reached what we call the {\it statistical limit}.
Otherwise, for timescales relatively short, the interaction is not
Markovian and it is very important to take into account the effects
of the interactions on the particle and its dynamics (memory effects).
The diffusion of a single adsorbate is thus modeled by a series of
random pulses within a Markovian regime (i.e., pulses of relatively
short duration in comparison with the system relaxation) simulating
collisions between adsorbates.
In particular, we describe these adsorbate-adsorbate collisions by
means of a {\it white shot noise} \cite{gardiner} as a limiting case
of colored shot noise.
In this way, a typical LMD problem involving $N$ adsorbates is
substituted by the dynamics of a single adsorbate, where the action
of the remaining $N-1$ adparticles is replaced by a random force
described by the white shot noise.
In this approximation, the distinction between self and distinct
time-dependent pair correlation function does not exist and
equations~(\ref{eq:DRP2}) and (\ref{eq:DSF}) still hold.
The intermediate scattering function reads now as
\be
 I(\Delta {\bf K},t) \equiv
  \langle e^{-i\Delta {\bf K} \cdot
   [{\bf R}(t) - {\bf R}(0)] } \rangle
  = \langle e^{-i\Delta {\bf K} \cdot
    \int_0^t {\bf v} (t') \ \! dt'} \rangle
 \label{eq:IntSF}
\ee
In order to get some analytical results and therefore a guide for the
interpretation of the numerical Langevin simulations, the intermediate
scattering function can be expressed as a second order cumulant
expansion in $\Delta {\bf K}$,
\begin{equation}
 I(\Delta {\bf K},t) \approx
  e^{- \Delta K^2 \int_0^t (t - t')
   \mathcal{C}_{\Delta {\bf K}}(t') dt'} ,
 \label{eq:IntSF2}
\end{equation}
where
\begin{equation}
 \mathcal{C}_{\Delta {\bf K}}(\tau) \equiv
 \langle v_{\Delta {\bf K}}(0) \ \!
  v_{\Delta {\bf K}}(\tau) \rangle \equiv
  \lim_{\mathcal{T}\to\infty} \frac{1}{\mathcal{T}}
  \int_0^\mathcal{T} v_{\Delta {\bf K}}(t) \ \!
   v_{\Delta {\bf K}}(t+\tau) \ \! dt
 \label{vcorr1}
\end{equation}
is the {\it autocorrelation function} of the velocity projected onto
the direction of the parallel momentum transfer (whose length is
$\Delta K \equiv \| \Delta {\bf K} \|$).
Only differences between two times, $\tau$, are considered because
this function is {\it stationary}.
This is the so-called {\it Gaussian approximation} \cite{mcquarrie},
which is exact when the velocity correlations at more than two
different times are negligible, thus allowing to replace the average
acting over the exponential function by an average acting over its
argument.
This approximation results of much help in the interpretation of
numerical simulations as well as in getting an insight into the
underlying dynamics.

The decay of $\mathcal{C}_{\Delta {\bf K}}(\tau)$ allows to define
a characteristic time, the {\it correlation time},
\begin{equation}
 \tilde{\tau} \equiv \frac{1}{\langle v_0^2 \rangle}
  \int_0^\infty \mathcal{C}_{\Delta {\bf K}}(\tau) \ \! d\tau ,
 \label{tauc}
\end{equation}
where $\sqrt{\langle v_0^2 \rangle} = \sqrt{k_B T/m}$ is the average
thermal velocity in one dimension ---though the dimensionality is
two, note that $\mathcal{C}_{\Delta {\bf K}}$ is defined along a
particular direction (that given by $\Delta {\bf K}$) and therefore
$\mathcal{C}_{\Delta {\bf K}}(0) \equiv k_B T/m$, i.e., the square of
the thermal velocity in one direction.
The correlation time is related to the line shape broadening in the
sense that it provides a timescale for the decay of the intermediate
scattering function and therefore information about the width of the
Q peak in the dynamic structure factor.


\subsubsection{Some elementary notions on Gaussian white noise}
\label{sec2.2.1}

Alternatively to the Einstein-Wiener stochastic model for Brownian
motion, in 1930 Orstein and Uhlenbeck formulated another one taking
the particle velocity as the stochastic variable of interest (rather
than its position, which is found integrating the velocity).
The starting point for this model is the equation proposed by Langevin
in 1908,
\begin{equation}
 m \dot{v} = - m \gamma v + m R_G (t) .
 \label{eq:3.1}
\end{equation}
This the simplest expression for an equation describing the Brownian
motion of a particle of mass $m$ in one dimension.
The right hand side (r.h.s.) of this equation can be split up into two
contributions: (1) a deterministic part characterized by the friction
force $- m \gamma v$ ($\gamma$ being the friction coefficient
depending on the fluid viscosity) and (2) a random part governed by
the random force $m R_G (t)$, where $R_G$ is a Gaussian white noise.
The random force (or, equivalently, the Gaussian white noise) satisfies
two conditions:
\begin{itemize}
\item[(i)]  The process $R_G (t)$ is Gaussian with zero mean, i.e.,
 $\langle R_G (t) \rangle = 0$.

\item[(ii)] The force-force correlation time is infinitely short,
 i.e., $m^2 \langle R_G (0) R_G (\tau) \rangle = A \delta(\tau)$,
 $A$ being a constant that gives the strength of the coupling
 between the adparticle and the environment.
\end{itemize}
The validity of this model relies on the fact that the Brownian
particle is much heavier than the particles constituting the
environment.
This implies that the kicks received by the particle of interest,
though relatively weak, are very effective when considered in a very
large number (the central limit theorem holds and therefore the noise
is Gaussian).
When applied to the motion of single adsorbates, the kicks come
from the surface fluctuations at a given temperature.
Note that the time evolution of the environmental degrees of
freedom is not taken into account because their correlations decay
faster than those of the particle (Markovian approximation), as
expressed by the condition (ii).

The relationship between the friction in the Langevin equation
and the fluctuations of the random force is given by {\it the
fluctuation-dissipation theorem} \cite{kubo}, which reads as
\begin{equation}
 \gamma (\omega)  =  \frac{m}{k_B T} \int_0^\infty
  \langle \delta R_G (0) \ \! \delta R_G (\tau) \rangle \ \!
  e^{-i \omega \tau} \ \! d\tau
 \label{eq:3.3}
\end{equation}
where
\begin{equation}
 \delta R_G (t) \equiv R_G (t) - \langle R_G (t) \rangle
 \label{eq:3.31}
\end{equation}
is the fluctuation due to the random noise function $R_G(t)$.
Making use of the properties (i) and (ii) for the Gaussian white
noise, the r.h.s.\ of equation~(\ref{eq:3.3}) becomes
\be
 \gamma (\omega) = \frac{A}{2 m k_B T} .
 \label{eq:3.4}
\ee
That is, the frequency spectrum of the friction force is flat, or
{\it white} in the sense that all frequencies contribute equally to
this spectrum, in analogy to white light.
This allows to establish
\begin{equation}
 \gamma(\omega) \equiv \gamma ,
 \label{eq:3.5}
\end{equation}
with the strength of the coupling between the Brownian particle and
the environment thus being
\begin{equation}
 A = 2 m \gamma k_B T ,
 \label{eq:3.7}
\end{equation}
The Gaussian white noise correlation function can now be defined as
\be
 \mathcal{G}_G (\tau) \equiv \langle \delta R_G (0)
  \ \! \delta R_G (\tau) \rangle
 = \frac{2 \gamma k_B T}{m} \ \! \delta (\tau) .
 \label{corrG2}
\ee
This dynamics implies that at thermal equilibrium the equipartition
theorem holds.


\subsubsection{Some elementary notions on shot noise}
\label{sec2.2.2}

The concept of noise arises from the early days of radio, the so-called
{\it shot noise} \cite{shot-noise} being one of the main sources of
noise, which was first considered by Schottky \cite{schottky} in 1918.
Studies on this type of noise were developed during the 1920's and
1930's, and were summarized and largely completed by Rice \cite{rice}
in the mid 1940's.
The paradigm of shot noise is the non steady electrical current
generated by independent (i.e., non correlated) electrons arriving at
the anode of a vacuum tube.
This (time-dependent) electric current can be expressed as
\begin{equation}
 I(t) = \sum_i b_i (t - t_i) ,
 \label{elect}
\end{equation}
where the pulse function $b_i(t-t_i)$ represents the contribution to
the current due to the $i$th individual electron, which is assumed to
be identical for each electron.
Moreover, it is also assumed that each electron arrives independently
of the previous ones.
The arrival times $t_i$ are thus randomly distributed with a certain
average number per unit time according to a Poisson distribution.

In our case, the current of electrons is replaced by the impacts
received by an adsorbate from other surrounding adsorbates.
We express this random force as $m \delta R_S(t)$, where
\begin{equation}
 \delta R_S(t) \equiv R_S(t) - \langle \langle R_S \rangle \rangle
 \label{rand-noise2}
\end{equation}
after making use of equation~(\ref{eq:3.31}), and where
\begin{equation}
 \langle \langle R_S \rangle \rangle \equiv
  \sum_K P_K (\mathcal{T}) \langle R_S(t') \rangle_\mathcal{T} .
 \label{doubleav}
\end{equation}
As seen, the double average bracket in the last expression indicates
averaging over the number of collisions ($K$) according to a certain
distribution ($P_K$) and the total time considered
($\mathcal{T}$).
In analogy to equation~(\ref{elect}) we have
\begin{equation}
 R_S(t) = \sum_{k=1}^K b_k (t - t_k) .
 \label{elect2}
\end{equation}
Here, $b_k(t-t_k)$ provides information about the shape and effective
duration of the $k$th adsorbate-adsorbate collision at $t_k$.
Then the probability to observe $K$ collisions after a time
$\mathcal{T}$ follows a Poisson distribution \cite{gardiner},
\begin{equation}
 P_K (\mathcal{T}) =
  \frac{\ (\lambda \mathcal{T})^K}{K!} \ \! e^{-\lambda\mathcal{T}} ,
 \label{poisson}
\end{equation}
where $\lambda$ is the average number of collisions per unit time.
That is, now the friction coefficient has to be interpreted in terms
of the collision frequency between adsorbates.

Assuming sudden adsorbate-adsorbate collisions (i.e., strong but
elastic collisions) and that after-collision effects relax
exponentially at a constant rate $\lambda'$, the pulses in
equation~(\ref{elect2}) can be modeled as
\begin{equation}
 b_k(t-t_k) = c_k \lambda' {\rm e}^{- \lambda' (t-t_k)} ,
 \label{pulse}
\end{equation}
with $t-t_k > 0$ and $c_k$ giving the intensity of the collision
impact.
Within a realistic model, collisions take place randomly at different
orientations and energies.
Hence it is reasonable to assume that the $c_k$ coefficients are
distributed according to an exponential law,
\begin{equation}
 g(c_k) = \frac{1}{\alpha} \ \! e^{-c_k/\alpha} , \qquad c_k \geq 0 ,
 \label{elaw}
\end{equation}
where the value of $\alpha$ will be determined later on.
Independently of their intensity, within this model any pulse decays
at the same rate $\lambda'$, as seen in equation~(\ref{pulse}).
This rate defines a decay timescale for collision events,
$\tau_c = 1/\lambda'$.
On the other hand, the (collision) friction coefficient introduces a
new timescale $\tau_r = 1/\lambda$, which can be interpreted as the
(average) time between two successive collisions. The diffusion motion
of the interacting adsorbates will take a time of the order of
$\tau_r$ in getting damped.

As previously done with the Gaussian white noise, we can now compute
the time correlation function of the shot noise
\begin{equation}
 \mathcal{G}_S(\tau) =
  \langle \langle \delta R_S (0)
   \ \! \delta R_S (\tau) \rangle \rangle ,
 \label{gtau1}
\end{equation}
where the double bracket is defined as in equation~(\ref{doubleav}).
A general expression for $\mathcal{G}_S (\tau)$ can be readily obtained
after straightforward algebraic manipulations, which yield
\begin{equation}
 \mathcal{G}_S (\tau) =
  \frac{1}{\mathcal{T}} \sum_K P_K (\mathcal{T}) K
  \int_0^\mathcal{T} \langle b(t - t') \ \! b(t + \tau - t') \rangle_c
   \ \! dt' ,
 \label{firstg}
\end{equation}
where
\begin{equation}
 \langle \ \cdot \ \rangle_c \equiv \int_0^\infty \cdot \ g(c) \ \! dc
\end{equation}
is the average over the pulse intensity, with $g(c)$ given by
equation~(\ref{elaw}).
Taking into account that $\sum_K P_K (\mathcal{T}) K =
\lambda \mathcal{T}$ and introducing the change of variable
$\zeta = t - t'$, equation~(\ref{firstg}) can be written in
an approximated way as
\begin{equation}
 \mathcal{G}_S (\tau) =
  \lambda \int_{-\infty}^\infty
   \langle b(\zeta) \ \! b(\zeta + \tau) \rangle_c \ \! d\zeta .
 \label{corrf}
\end{equation}
This approximation relies on the hypothesis that $b(\zeta) \approx 0$
outside a narrow time interval, $0 < \zeta < \Delta$, with $\Delta$
a few times larger than $\tau_c$, but of the same order of magnitude.
Thus, equation (\ref{corrf}) is a general expression independent of
the shape considered to simulate the pulses.
Substituting equation~(\ref{pulse}) into (\ref{corrf}) one then
obtains
\begin{equation}
 \mathcal{G}_S (\tau) =
  \frac{\lambda \lambda'}{\alpha^2} \ \! {\rm e}^{- \lambda' |\tau|} .
 \label{corrfe}
\end{equation}

The validity of the standard Langevin approach to study a dynamics
governed by a shot noise is determined by the fluctuation-dissipation
theorem \cite{kubo}.
As will be seen, in the generalized Langevin formulation we have a
memory function in terms of the time-correlation function instead of
a (time-independent) friction coefficient.
According to the fluctuation-dissipation theorem, the formal
relationship between the frequency spectrum of the memory function
and the random force correlation function is given by

\begin{equation}
 \tilde{\xi} (\omega) = \frac{m}{k_B T}
  \int_0^\infty \mathcal{G}_S (\tau)
   \ \! e^{- i \omega \tau} \ \! d\tau .
 \label{fdt}
\end{equation}
Introducing equation~(\ref{corrfe}) into this integral yields
\begin{equation}
 \tilde{\xi} (\omega) =
  \lambda \ \! \frac{\lambda'}{\lambda' + i \omega} ,
 \label{fdte}
\end{equation}
whose real part is
\begin{equation}
 {\rm Re} [ \tilde{\xi} (\omega) ]
  = \frac{1}{2} \ \! [ \tilde{\xi} (\omega) + \tilde{\xi}^* (\omega) ]
  = \lambda \ \! \frac{\lambda'^2}{\lambda'^2 + \omega^2} .
 \label{rfdte}
\end{equation}
Two limits are interesting in this expression: $\lambda' \ll \omega$
and $\lambda' \gg \omega$.
The first limit involves very short timescales (smaller than $\tau_c$).
Memory effects are important and the generalized Langevin equation
should be applied.
Note that in this case, equation~(\ref{fdte}) can be written as
\begin{equation}
 \tilde{\xi} (\omega) \approx \lambda \ \! \frac{\lambda'^2}{\omega^2} ,
 \label{limit1}
\end{equation}
and this frequency-dependent friction does not allow to define an
appropriate relaxation timescale $\tau_r$ (colored shot noise).
Conversely, in the second case, the collision timescale rules the
system dynamics; it establishes a cutoff frequency, which leads to
\begin{equation}
 \tilde{\xi} (\omega) \approx \lambda
  \left( 1 - \frac{\omega^2}{\lambda'^2} \right) .
 \label{gammaapprox}
\end{equation}
This expression can be approximate as $\tilde{\xi} (\omega) \approx
\lambda$ whenever $\lambda \ll \omega \ll \omega_c = \tau_c^{-1}$
(i.e., $\tau_r \ll \tau_c$).
This limit holds for strong but localized (or instantaneous) collisions
(as assumed here) as well as for weak but continuous kicks (Brownian
motion).
Since it is similar to the condition leading to
equation~(\ref{eq:3.5}), we can speak about a Poissonian white noise
and make use of the standard Langevin equation.

\begin{figure}
 \begin{center}
 \epsfxsize=3.5in {\epsfbox{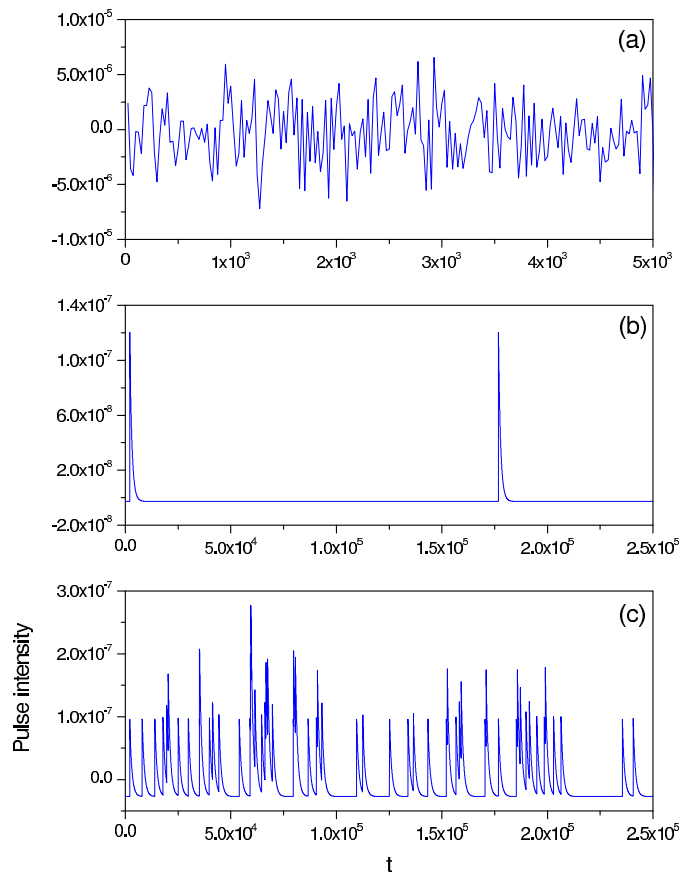}}
 \caption{\label{fig1}
  (a) Random noise function, $\delta R_G (t)$,
  corresponding to a Gaussian white noise with $\gamma = 0.1 \ \!
  \omega_0$ ($\omega_0 = 2.2049\times10^{-4}$ a.u.\ is the harmonic
  frequency associated with the nonseparable adsorbate-substrate
  interaction potential (see section~\ref{sec3.1})).
  (b) Random noise function, $\delta R_S (t)$, corresponding to a
  shot noise with $\lambda = \gamma$.
  (c) Random noise function, $\delta R_S (t)$, corresponding to a
  shot noise with $\lambda = 10 \ \! \gamma$.
  The calculations refer to Na atoms at $T = 200$~K and assuming
  $\lambda' = 10^{-3}$ a.u.\ (41.3~ps~$^{-1}$).}
 \end{center}
\end{figure}

Due to their influence in the diffusion process, it is worth comparing
the time series corresponding to $\delta R_G (t)$ and $\delta R_S (t)$.
For the sake of simplicity, we consider that all pulses in the shot
noise have the same intensity ---it can be easily shown that this is
equivalent to assume the pulse intensities distributed according to
equation~(\ref{elaw}), though substituting $\langle c_k \rangle$ and
$\langle c_k^2 \rangle$ by $C$ and $C^2$ ($C = 2k_B T/m$ being the
impact strength), respectively, wherever such averages appear.
In figure~\ref{fig1}(a) we observe the time series corresponding to a
Gaussian white noise, $\delta R_G (t)$.
This series presents a fractal-like profile, i.e., enlargements of
any subinterval will look like the whole interval considered.
This is because within any relatively short period of time the particle
feels many (uncorrelated) kicks from the surface in both directions,
positive and negative.
This is not the case, however, for a shot noise with $\lambda =
\gamma$.
As seen in figure~\ref{fig1}(b), the particle receives 2 kicks in
average from another adsorbates in a much larger period of time
(about 50 times the one considered for $\delta R_G (t)$).
This means that in principle the mean free path for a single
adparticle is relatively large.
However, though a single adparticle receives very few hits, the
collective effect is similar to that of having a Gaussian white noise
because of the random distributions of the kicks.
The effect is more apparent when larger values of $\lambda$ are
considered, because the mean time between consecutive collisions
decreases dramatically, as can be seen in figure~\ref{fig1}(c).
In any case, both noises are completely uncorrelated, that is,
\be
 \langle \delta R_G (t) \delta R_S (t') \rangle = 0 .
 \label{prop3}
\ee

\subsubsection{Relationship between coverage and the collision frequency}
\label{sec2.2.3}

Here we would like to show how the coverage $\theta$ and $\lambda$ can
be related in a simple manner.
In the elementary kinetic theory of transport in gases \cite{mcquarrie}
diffusion is proportional to the mean free path $\bar{l}$, which is
proportionally inverse to both the density of gas particles and the
effective area of collision when a hard-sphere model is assumed.
For two-dimensional collisions the effective area is replaced by an
effective length (twice the radius $\rho$ of the adparticle) and the
gas density by the surface density $\sigma$.
Accordingly, the mean free path is given by
\begin{equation}
 \bar{l} = \frac{1}{2 \sqrt{2} \rho \sigma} .
 \label{mfp}
\end{equation}
Taking into account the Chapman-Enskog theory for hard spheres, the
self-diffusion coefficient can be written as
\begin{equation}
 D = \frac{1}{6 \rho \sigma} \ \! \sqrt{\frac{k_B T}{m}} .
 \label{d-mfp}
\end{equation}
Now, from Einstein's relation (see section~\ref{sec2.2.4}), and taking
into account that $\theta = a^2 \sigma$ for a square surface lattice
of unit cell length $a$, we finally obtain
\begin{equation}
 \lambda = \frac{6\rho}{a^2} \ \! \sqrt{\frac{k_B T}{m}} \ \! \theta .
 \label{theta}
\end{equation}
Therefore, given a certain surface coverage and temperature, $\lambda$
can be readily estimated from equation~(\ref{theta}).
Notice that when the coverage is increased by one order of magnitude,
the same holds for $\lambda$ at a given temperature.


\subsubsection{Interacting adsorbate dynamics within the Langevin
 equation}
\label{sec2.2.4}

Except for the case of a free-potential regime, the analytical study
of particle motion in two dimensions results intractable in general
due to the correlations between the two degrees of freedom.
However, since we are interested in offering an analytical formulation
that allows to better understand the process ruling surface diffusion
and low vibrational motions, it is sufficient to proceed in one
dimension and then try to adapt the resulting formulation to two
dimensions.
Thus, the motion of an adsorbate subjected to the action of a bath
consisting of another adsorbates on a static one-dimensional surface
potential can be well described by a generalized Langevin equation,
\begin{equation}
 \ddot{x}(t) = - \int_0^t \eta (t-t') \ \! \dot{x}(t') \ \! dt'
  + F(x(t)) + \delta R(t) ,
 \label{eq-langg}
\end{equation}
where $x$ represents the adsorbate coordinate, and its first and second
time derivatives are expressed by one and two dots on the $x$
variable, respectively.
In this equation, $\eta(t)$ is the bath memory function, which includes
the effects arising from both the Gaussian white noise and the shot
noise. Thus, the noise source is expressed as the sum of both
contributions,
\begin{equation}
 \delta R_{GS} (t) = \delta R_G (t) + \delta R_S (t) .
\end{equation}
Regarding the deterministic term, $F = - \nabla V$ is the deterministic
force per mass unit derived from the periodic surface interaction
potential ($V(x) = V(x+a)$, $a$ being the period along the $x$
direction).
If $\tau_c$ is relatively small (i.e., collision effects relax
relatively fast), the memory function in equation~(\ref{eq-langg})
will be local in time.
This allows to express the memory function as $\eta (t-t') \simeq
(\gamma + \lambda) \delta(t-t')$ and expand the upper time limit in
the integral to infinity. Taking into account this approximation
and from the fluctuation-dissipation theorem, which leads to $\eta =
\gamma + \lambda$, equation~(\ref{eq-langg}) becomes
\begin{equation}
 \ddot{x}(t) = - \eta \dot{x}(t) + F(x(t))
  + \delta R_G(t)+ \delta R_S(t) .
 \label{eq-lang1}
\end{equation}
The solutions of this equation can be straightforwardly obtained by
formal integration, this rendering
\begin{eqnarray}
 v(t) & = & v_0 e^{- \eta t}
   + \int_0^t e^{- \eta (t-t')} F(x(t')) \ \! dt'
   + \int_0^t e^{- \eta (t-t')} \delta R_{GS} (t') \ \! dt' ,
 \label{veloc1} \\
 x(t) & = & x_0 + \frac{v_0}{\eta} \ \! ( 1 - e^{- \eta t} )
   + \frac{1}{\eta} \int_0^t \Big[ 1 - e^{- \eta (t-t')} \Big]
    F(x(t')) \ \! dt'
  \nonumber \\
  & + & \frac{1}{\eta} \int_0^t \Big[ 1 - e^{- \eta (t-t')} \Big]
    \delta R_{GS} (t') \ \! dt' ,
 \label{posit12}
\end{eqnarray}
where $v_0 = v(0)$ and $x_0 = x(0)$.
As can be seen, for $\delta R_{GS} = 0$, equations~(\ref{veloc1}) and
(\ref{posit12}) become the formal solutions of purely deterministic
equations of motion.
Therefore, without loss of generality, these solutions can be more
conveniently expressed as
\begin{eqnarray}
 v(t) & = & v_d(t) + v_{s,G}(t) + v_{s,S}(t) ,
 \label{veloc2} \\
 x(t) & = & x_d(t) + x_{s,G}(t) + x_{s,S}(t) ,
 \label{posit2}
\end{eqnarray}
where $d$ refers to the deterministic terms and $s$ to those depending
on the stochastic forces.
Nonetheless, note that when $\delta R_{GS} (t) \ne 0$ the deterministic
part will also present some stochastic features due to the evaluation
of $F(x)$ along the trajectory $x(t)$, which is stochastic.

Taking advantage of equation~(\ref{prop3}), we can write
\begin{eqnarray}
 \langle v(t) \rangle & = & \bar{v}_d(t) ,
 \label{avveloc12} \\
 \langle v^2(t) \rangle & = & \bar{v}^2_d(t)
  + \langle v_{s,G}^2(t) \rangle +\langle v_{s,S}^2(t) \rangle  ,
 \label{avveloc22} \\
 \langle x(t) \rangle & = & \bar{x}_d(t) ,
 \label{avposit12} \\
 \langle x^2(t) \rangle & = & \bar{x}^2_d(t)
  + \langle x_{s,G}^2(t) \rangle+\langle x_{s,S}^2(t) \rangle  ,
 \label{avposit22}
\end{eqnarray}
where the barred magnitudes indicate the respective averages of the
deterministic part of the solutions, and
\begin{eqnarray}
 \langle v_{s,X}^2(t) \rangle & = &
  e^{- 2 \eta t} \int_0^t dt' \ \! e^{2 \eta t'}
   \int_{-t'}^{t-t'} e^{\eta \tau} \mathcal{G}_X (\tau) \ \! d\tau ,
 \label{stocveloc1} \\
 \langle x_{s,X}^2(t) \rangle & = &
   \frac{1}{\eta^2}
   \int_0^t dt' \ \! \Big[ 1 - e^{- \eta (t-t')} \Big]
   \int_{-t'}^{t-t'} \Big[ 1 - e^{- \eta (t-t'-\tau)} \Big] \ \!
    \mathcal{G}_X (\tau) \ \! d\tau ,
 \label{stocposit1}
\end{eqnarray}
with $X = G$ or $S$.
For $X = G$, the final form of these expressions reads as
\begin{eqnarray}
 \langle v_{s,G}^2(t) \rangle & = & \frac{\gamma}{\eta}
  \frac{1}{\alpha^2} \ \! \Big( 1 - e^{- 2 \eta t} \Big)
 \label{stocvelocg} \\
 \langle x_{s,G}^2(t) \rangle & = & \frac{\gamma}{\eta^3}
  \frac{1}{\alpha^2} \ \! \Big[ 2 \eta t + 1
   - \Big( 2 - e^{- \eta t} \Big)^2 \Big] ,
 \label{stocpositg}
\end{eqnarray}
and for $X = S$ as
\begin{eqnarray}
 \langle v_{s,S}^2(t) \rangle & = & \frac{1}{\alpha^2} \ \! \bigg\{
  \frac{\lambda}{\eta} \frac{\lambda'}{(\lambda' - \eta)}
   \ \! \Big( 1 - e^{- 2 \eta t} \Big)
  - \frac{2 \lambda' \lambda}{\lambda'^2 - \eta^2} \ \!
    \Big[ 1 - e^{- (\lambda' + \eta) t} \Big] \bigg\} ,
 \label{stocvelocs} \\
 \langle x_{s,S}^2(t) \rangle & = & \frac{\lambda}{\eta^3}
  \frac{1}{\alpha^2} \ \! \bigg\{ 2 \eta t
  - \frac{4 \lambda' - 2 \eta}{\lambda' - \eta} \ \!
     \Big( 1 - e^{- \eta t} \Big)
  + \frac{2 \lambda'}{2 (\lambda' - \eta)} \ \!
   \Big( 1 - e^{- 2 \eta t} \Big)
 \nonumber \\
  & + & \frac{2 \eta^2}{\lambda' (\lambda' - \eta)} \ \!
        \Big( 1 - e^{- \lambda' t} \Big)
    - \frac{2 \eta^2}{\lambda'^2 - \eta^2} \ \!
      \Big[ 1 - e^{- (\lambda' + \eta) t} \Big] \bigg\} .
 \label{stocposits}
\end{eqnarray}
Introducing now the assumption $\lambda' \gg \lambda$, we obtain
\begin{eqnarray}
 \langle v_{s,S}^2(t) \rangle & \approx & \frac{\lambda}{\eta}
  \frac{1}{\alpha^2} \ \! \Big( 1 - e^{- 2 \eta t} \Big) ,
 \label{stocvelocs2} \\
 \langle x_{s,S}^2(t) \rangle & \approx & \frac{\lambda}{\eta^3}
  \frac{1}{\alpha^2} \Big[ 2 \eta t + 1
   - \Big( 2 - e^{- \eta t} \Big)^2 \Big] .
 \label{stocposits2}
\end{eqnarray}
These equations are identical to equations~(\ref{stocvelocg}) and
(\ref{stocpositg}), except for a weighting factor $\lambda/\eta$.
The weighting factors $\gamma/\eta$ and $\lambda/\eta$ indicate
the contribution arising from each noise.
Adding equations (\ref{stocvelocg}) and (\ref{stocvelocs2}), on the
one hand, and equation~(\ref{stocpositg}) and (\ref{stocposits2}),
on the other hand, we reach
\begin{eqnarray}
 \langle v_s^2(t) \rangle & \approx &
  \frac{1}{\alpha^2} \ \! \Big( 1 - e^{- 2 \eta t} \Big) ,
 \label{stocvelf} \\
 \langle x_s^2(t) \rangle & \approx & \frac{1}{\eta^2}
  \frac{1}{\alpha^2} \Big[ 2 \eta t + 1
   - \Big( 2 - e^{- \eta t} \Big)^2 \Big] ,
 \label{stocposf}
\end{eqnarray}
where no clue about the type of noise is left.
Equation~(\ref{stocvelf}) can now be used to determine the value
of $\alpha$; assuming that the equipartition theorem satisfies for
$t \to \infty$,
\begin{equation}
 \frac{1}{2} \ \! m \langle v^2(\infty) \rangle =
  \frac{1}{2} \ \! k_B T .
\end{equation}
Therefore, since $\bar{v}^2_d(t) = \bar{v}^2_0 e^{-2\eta t}$ and
the time-dependent term in equation~(\ref{stocvelf}) vanishes
asymptotically,
\begin{equation}
 \alpha = \sqrt{\frac{m}{k_B T}} .
 \label{value}
\end{equation}
If the system is initially thermalized (i.e., it follows a
Maxwell-Boltzmann distribution in velocities) and has a uniform
probability distribution in positions around $x=0$, then
$\bar{v}_0 = 0$, $\bar{v}^2_0 = k_B T / m$ and $\bar{x}_0 = 0$.
Thus, for $\lambda' \gg \lambda$ (i.e., in the limit of a Poissonian
white noise), we finally obtain
\begin{eqnarray}
 \langle v(t) \rangle & = & 0 ,
 \label{avveloc13} \\
 \langle v^2(t) \rangle & = & \frac{k_B T}{m} ,
 \label{avveloc23} \\
 \langle x(t) \rangle & = & 0 ,
 \label{avposit13} \\
 \langle x^2(t) \rangle & = & \bar{x}_0^2 +
  \frac{k_B T}{m \eta^2} \ \!
  \Big[ 2 \eta t + 1 - \Big( 2 - e^{- \eta t} \Big)^2 \Big] .
 \label{avposit23}
\end{eqnarray}
Obviously, these equations constitute a limit and therefore for values
of the parameters out of the range of the approximation deviations will
be apparent.

As happens with Brownian motion, two regimes are clearly
distinguishable from equation~(\ref{avposit23}).
For $\eta t \ll 1$ collision events are rare and the adparticle shows
an almost free motion with relatively long mean free paths.
This is the {\it ballistic} or {\it free-diffusion regime},
characterized by
\begin{equation}
 \langle x^2(t) \rangle \sim \frac{k_B T}{m} \ \! t^2 .
 \label{avvalues4}
\end{equation}
On the other hand, for $\eta t \gg 1$ there is no free diffusion
since the effects of the stochastic force (collisions) are dominant.
This is the {\it diffusive regime}, where mean square displacements
are linear with time:
\begin{equation}
 \langle x^2(t) \rangle \sim
  \frac{2 k_B T}{m \eta} \ \! t = 2 D t \ \! .
 \label{avvalues5}
\end{equation}
This is the so-called {\it Einstein's law}.
Note from equation~(\ref{avvalues5}) that: (1) lowering the friction
$\eta$ acting on the adparticle leads to a faster diffusion (the
diffusion coefficient $D$ increases) and (2) diffusion becomes more
active when the surface temperature increases.

\begin{figure}
 \begin{center}
 \epsfxsize=4in {\epsfbox{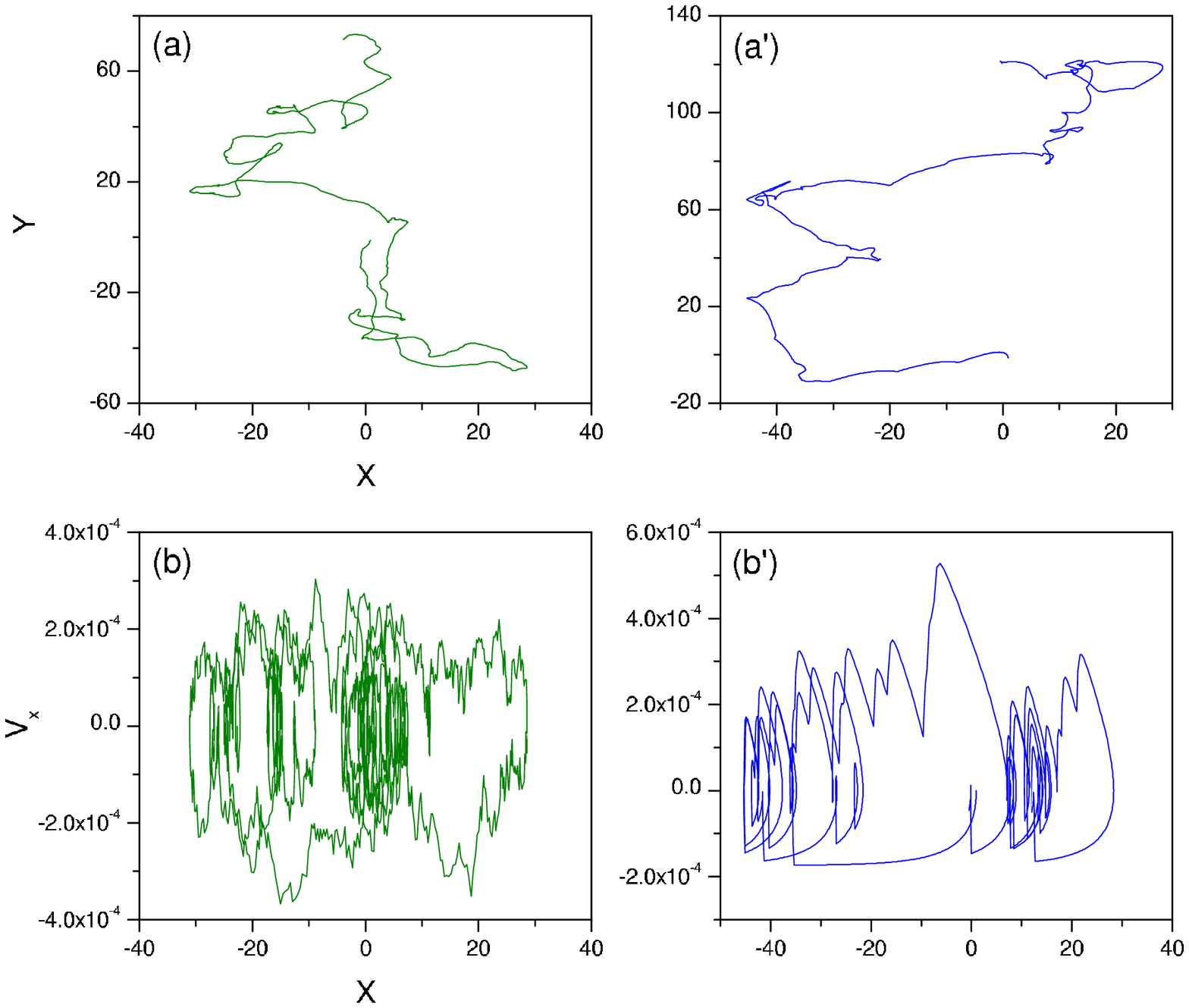}}
 \caption{\label{fig2}
  Upper panels: Trajectory dynamics for
  $V(x,y) = 0$ and: (a) a Gaussian white noise with $\gamma =
  0.1 \ \! \omega_0$ ($\omega_0 = 2.2049\times10^{-4}$ a.u.)
  and (a') a shot noise with $\lambda = \gamma$.
  Lower panels: Phase diagram corresponding to the trajectories
  represented above.
  The trajectories have been propagated for 2.5$\times$10$^6$~a.u.\
  and $\lambda' = 10^{-3}$ a.u.}
 \end{center}
\end{figure}

\begin{figure}
 \begin{center}
 \epsfxsize=4in {\epsfbox{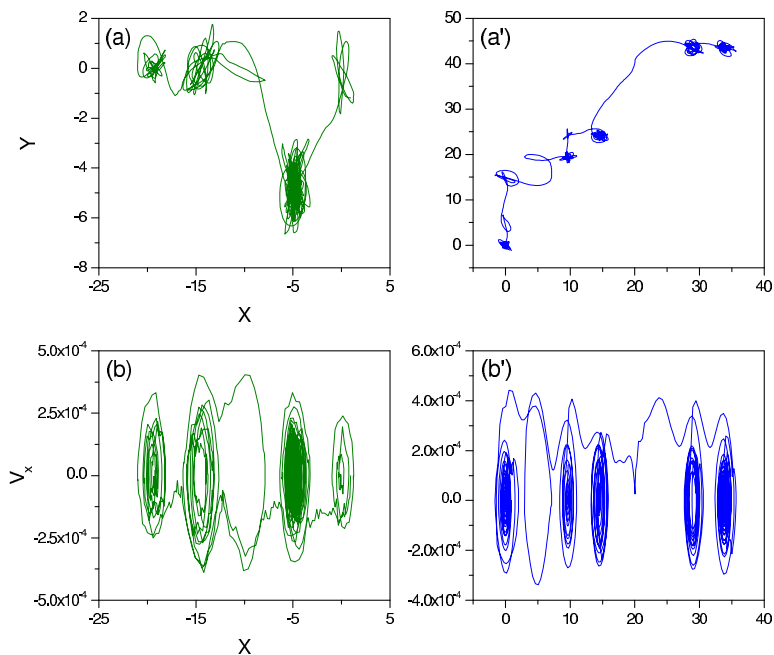}}
 \caption{\label{fig3}
  Upper panels: Trajectory dynamics for
  $V(x,y)$ as given by model~3 (see section~\ref{sec3.1}) and:
  (a) a Gaussian white noise with $\gamma = 0.1 \ \! \omega_0$
  and (a') a shot noise with $\lambda = \gamma$.
  Lower panels: Phase diagram corresponding to the trajectories
  represented above.
  The trajectories have been propagated for 2.5$\times$10$^6$~a.u.\
  and $\lambda' = 10^{-3}$ a.u.}
 \end{center}
\end{figure}

To conclude this section, it is worth comparing the type of dynamics
ruled by each noise, Gaussian or shot.
In figures~\ref{fig2} and \ref{fig3} we present some trajectories
together with their phase diagram (for the $x$-coordinate) for $V(x,y)
= 0$ and $V(x,y)$ as given by equation~(\ref{eq:Pot0}) (see
section~\ref{sec3}), respectively.
The noise functions are such that $\lambda = \gamma$
(figures~\ref{fig1}(a) and \ref{fig1}(b)) and the numerical details
are the same as in figure~\ref{fig1} (the total propagation time is
$t = 2.5\times10^6$~a.u.).
The first striking difference that we can appreciate is that, because
of the continuous kicking in the case of a Gaussian white noise, the
trajectory looks much smoother in the case of a shot noise.
This can be better seen in the upper panels of figure~\ref{fig3} ---the
effect is not so pronounced in figure~\ref{fig2} because we have only
represented one out of 50 time steps--- and also in the lower panels of
both figures.
The lack of smoothness in the trajectories affected by the Gaussian
white noise is a manifestation of this type of kicking (see
figure~\ref{fig1}(a)).
On the other hand, it is also interesting to observe the difference
in the dynamics induced in both cases when a potential is introduced.
The presence of potential wells, where a particle can
remain trapped, will make the particle to undergo a vibrational
motion inside such wells. The low frequency vibrational motion observed
gives rise to the so-called {\it frustrated translational mode}
or T {\it mode}.

\subsubsection{The velocity autocorrelation function}
\label{sec2.2.5}

The average magnitudes given in section~\ref{sec2.2.4} are only
relevant to understand the ensemble behaviour of the trajectories along
time, and therefore to have some idea on the dynamics induced by the
corresponding interaction potential.
In order to get a deeper insight into the diffusion process ---and
therefore also into the line shapes associated with the dynamic
structure factor--- it is important first to compute the velocity
autocorrelation function given by equation~(\ref{vcorr1}).
This function can be expressed as
\begin{equation}
 \mathcal{C}(\tau) = \langle v_{\Delta {\bf K}}(t) \ \!
  v_{\Delta {\bf K}}(t+\tau) \rangle
  = e^{- 2 \eta t - \eta \tau} \int_0^t dt' \ \! e^{2 \eta t'}
   \int_{-t'}^{t+\tau-t'} e^{\eta s} \mathcal{G}_{GS} (s) \ \! ds .
 \label{vcorr2}
\end{equation}
As seen, this expression only depends on the stochastic part of the
solution for the velocity; the deterministic one cancels out in the
averaging procedure.
Moreover, it also takes into account that there are no correlations
among the two space coordinates ($x$ and $y$).
This is true in any case for the Gaussian white noise; for the shot
noise it only holds in the white noise limit, otherwise some
correlations among the two degrees of freedom could still be present
when $V \neq 0$ due to the finite duration of the pulses.
To avoid this inconvenience, which deviates the treatment from
analyticity, here we assume conditions for which the white noise limit
is satisfied (see section~\ref{sec2.2.4}).
Finally, note that in equation~(\ref{vcorr2}) it is also considered
that the motion along both coordinates contribute equally (i.e.,
$\langle v_x(0)\ \! v_x(t)\rangle \approx \langle v_y(0)
\ \! v_y(t) \rangle$).
This finds its justification in the fact that the potential models
that will be used display the same symmetry along each direction and
therefore the corresponding dynamics is expected to be the same (this
would not be the case, for example, when the activation barrier is much
higher in one direction).
In this sense, the calculation of the velocity autocorrelation function
can be regarded as one-dimensional.

The analytical results derived following the assumptions described
above will be compared with the numerical simulations, which have been
obtained using the corresponding exact expressions (e.g., the r.h.s.\
of the first equality in equation~(\ref{eq:IntSF})).
In order to better interpret our exact numerical results two limiting
cases will be studied below.
It is precisely the discussion in terms of both cases what will allow
us to better understand the physical processes taking place on the
surface.


\subsubsection{The low corrugation regime. Quasi-free adparticles.}
\label{sec2.2.6}

In the case of diffusion on low corrugated surfaces, where the role of
the adiabatic adsorbate-substrate interaction potential is negligible
and only the action of the thermal phonons is relevant, one can assume
$V \approx 0$.
The adparticle motion can then be regarded as quasi-free since it is
not ruled by a potential, but only influenced by the stochastic forces.
Within this regime, equation~(\ref{vcorr2}) becomes
\begin{equation}
 \mathcal{C}(\tau) =
   \frac{\gamma}{\eta} \frac{k_B T}{m} \ \! e^{- \eta \tau}
 + \frac{\lambda}{\eta} \frac{k_B T}{m}
   \frac{\lambda'^2}{\lambda'^2 - \eta^2}
    \left( e^{- \eta \tau} -
      \frac{\eta}{\lambda'} \ \! e^{- \lambda' \tau} \right) .
 \label{corrGM1}
\end{equation}
In this expression the contributions arising from each noise source are
apparent.
However, in the limit of Poissonian white noise ($\lambda' \gg\lambda$)
such a distinction disappears; under such a condition
equation~(\ref{corrGM1}) becomes
\begin{equation}
 \mathcal{C}(\tau) = \frac{k_B T}{m} \ \! e^{- \eta \tau} .
 \label{corrGM2}
\end{equation}
Unless the relaxation of the collisional effects is relevant, these
effects and those caused by the surface should be indistinguishable
and equation~(\ref{corrGM2}) would describe accurately the loss of
correlation, with $\tilde{\tau}$ given in terms of $\eta$ rather than
$\gamma$ and/or $\lambda$ separately.

The expression for the intermediate scattering function resulting
from equation~(\ref{corrGM2}) is
\begin{equation}
 I(\Delta K,t) = \exp \left[- \chi^2
   \left( e^{- \eta t} + \eta t - 1 \right) \right] ,
 \label{eq:IntSGM}
\end{equation}
where the so-called {\it shape parameter} $\chi$
\cite{prb-I,jcp-elliot} is defined as
\begin{equation}
 \chi^2 \equiv \langle v_0^2 \rangle \Delta K^2 / \eta^2 .
 \label{eq:chi2}
\end{equation}
From this relation we can obtain both the mean free path $\bar{l}
\equiv \tilde{\tau} \sqrt{ \langle v_0^2 \rangle }$ and the
self-diffusion coefficient $D\equiv\tilde{\tau}\langle v_0^2\rangle$
({\it Einstein's relation}).
When the coverage increases, the collisions among adsorbates also
increase, and so $\lambda$ (see section~\ref{sec2.2.4}) and therefore
$\eta$.
As can be easily shown, equation~(\ref{eq:IntSGM}) displays a Gaussian
decay at short times that does not depend on the particular value of
$\eta$, while at longer times it decays exponentially with a rate given
by $\eta^{-1}$. Thus, with $\eta$, the decay of the intermediate
scattering function becomes slower.
The type of decay is important concerning the width and the shape of
the dynamic structure factor \cite{JLvega1,prb-I}.

The above described effects can be quantitatively understood by means
of the expression of the dynamic structure factor obtained analytically
from equation~(\ref{eq:IntSGM}),
\ba
 S (\Delta K, \omega) & = & \frac{e^{\chi^2}}{\pi \eta} \ \!
  \chi^{-2\chi^2} \ \! {\rm Re} \left\{ \chi^{-2i\omega/\eta}
   [ \tilde {\Gamma} (\chi^2 + i\omega/\eta)
    - \tilde {\Gamma} (\chi^2 + i\omega/\eta, \chi^2)] \right\}
 \nonumber \\
 & = & \frac{e^{\chi^2}}{\pi}
  \sum_{n=0}^\infty \frac{(-1)^n \chi^{2n}}{n!}
   \frac{(\chi^2 + n) \eta}{\omega^2 + [ (\chi^2 + n) \eta]^2} \ \! .
 \label{dsf1}
\ea
Here, the $\tilde{\Gamma}$ symbol in the first line denotes both the
Gamma and incomplete Gamma functions (depending on the corresponding
argument), respectively.
As can be noted in the high friction limit, equation~(\ref{dsf1})
becomes a Lorentzian function, its {\it full width at half maximum}
(FWHM) being $\Gamma = 2\eta\chi^2$, which approaches zero as $\eta$
increases (narrowing effect). This in sharp contrast to what one
could expect --- as the frequency between successive collisions increases
one would expect that the line shape gets broader (effect of the
pressure in the spectral lines of gases).
The physical reason for this effect could be explained as follows.
As $\eta$ increases the particle's mean free path decrease and
therefore correlations are lost more slowly.
In the limit case where friction is such that the particle remains in
the same place, the van Hove function becomes a $\delta$-function, the
intermediate scattering function remains equal to 1 and the dynamic
structure factor consists of a $\delta$-function at $\omega = 0$.
Conversely, in the low friction limit the line shape is given by a
Gaussian function, whose width is $\Gamma = 2\sqrt{2\ln 2}
\sqrt{k_B T/m} \ \! \Delta K$, which does not depend on $\eta$.
This is the case for a two--dimensional {\it free gas} \cite{prb-I,toeprl}.
This gradual change of the line shapes as a function of the friction
and/or the parallel momentum transfer leading to a change of the shape
parameter $\chi$ is known as the
{\it motional narrowing effect} \cite{JLvega0,JLvega1,prb-I}.
Note that in our approach the friction is related to the coverage.
Thus, at higher coverages a narrowing effect is predicted for a flat
surface \cite{prl-I}.


\subsubsection{The harmonic oscillator. Bound adparticles.}
\label{sec2.2.7}

In contrast with the case of a dynamics where $V$ does not play a
relevant role, we can devise a particle fully trapped within a
potential well.
The harmonic oscillator is an appropriate working model to understand
the physics associated with this problem.
This model also allows us to understand the behaviour associated with
the T mode, which comes precisely from the oscillating behaviour
undergone by the particle when the diffusional motion is temporarily
frustrated.

For a harmonic oscillator, the behaviour of the adparticle becomes very
apparent when looking at the corresponding velocity autocorrelation
function, which reads \cite{JLvega1,risken} as
\begin{equation}
 \mathcal{C}(\tau) = \frac{k_B T}{m} \ \! e^{- \eta \tau/2}
  \left( \cos \bar{\omega} \tau
   - \frac{\eta}{2 \bar{\omega}} \sin \bar{\omega} \tau \right) .
 \label{corrHO}
\end{equation}
Here,
\begin{equation}
 \bar{\omega} \equiv \sqrt{ \omega_0^2 - \frac{\eta^2}{4} } ,
\end{equation}
$\omega_0$ being the harmonic frequency.
Alternatively, equation~(\ref{corrHO}) can also be recast as
\begin{equation}
 \mathcal{C}(\tau) =
  \frac{k_B T}{m} \ \! \frac{\omega_0}{\bar{\omega}} \ \!
   e^{- \eta \tau/2} \cos (\bar{\omega} \tau + \delta) ,
 \label{corrHO2}
\end{equation}
with
\begin{equation}
 \delta \equiv (\tan)^{-1} \left( \frac{\eta/2}{\bar{\omega}} \right) .
\end{equation}
Note that equation~(\ref{corrGM2}) can be easily recovered after
some algebra in the limit $\omega_0 \to 0$ from either
equation~(\ref{corrHO}) or (\ref{corrHO2}).
In the case of anharmonic potentials, provided that we work within an
approximate harmonic regime, $\omega_0$ would represent the
corresponding approximate harmonic frequency.

The only information about the structure of the lattice is found
in the shape parameter through $\Delta K$ [see
equation~(\ref{eq:chi2})].
When large parallel momentum transfers are considered, both the
periodicity and the structure of the surface have to be taken into
account.
Consequently, the shape parameter should be changed for different
lattices.
The simplest model including the periodicity of the surface is
that developed by Chudley and Elliott \cite{elliott}, who proposed
a master equation for the pair-distribution function in space and
time assuming instantaneous discrete jumps on a two-dimensional
Bravais lattice. Very recently, a generalized shape parameter based on
that model has been proposed to be \cite{jcp-elliot}
\be
 \chi_l (\Delta {\bf K}) \equiv \sqrt{\frac{\Gamma_{\nu}
 (\Delta {\bf K})}{2 \eta}} ,
 \label{eq11}
\ee
where, for our approach, it has been written $\eta$ instead of
$\gamma$. Here, $\Gamma_{\nu} (\Delta {\bf K})$ represents the inverse
of the correlation time and is expressed as

\be
 \Gamma_{\nu} (\Delta {\bf K}) = \nu
 \sum_{\bf j} P_{\bf j} \ \! [1 - \cos({\bf j} \cdot \Delta {\bf K})] ,
 \label{eq9}
\ee
$\nu$ being the total jump rate out of an adsorption site and
$P_{\bf j}$ the relative probability that a jump with a displacement
vector ${\bf j}$ occurs.

Introducing now equation~(\ref{corrHO2}) into (\ref{eq:IntSF2}) leads
to the following expression for the intermediate scattering function
\begin{equation}
 I(\Delta K,t) =
  \exp \left\{ - \frac{\chi_l^2 \eta^2}{\bar{\omega} \omega_0}
   \left[ \cos \delta - e^{-\eta t/2}
    \cos (\bar{\omega} t - \delta) \right] \right\} .
 \label{isfho}
\end{equation}
The argument of this function displays an oscillatory behaviour around
a certain value with the amplitude of the oscillations being
exponentially damped.
This translates into an also decreasing behaviour of the intermediate
scattering function, which also displays oscillations around the
asymptotic value.
This means that after relaxation the intermediate scattering function
has not fully decayed to zero unlike the free-potential case.
Again, in the limit $\omega_0 \to 0$, equation~(\ref{isfho}) approaches
equation~(\ref{eq:IntSGM}).

In order to obtain an analytical  expression for the dynamic structure
factor, it is convenient to express equation~(\ref{isfho}) as
\ba
 I(\Delta K, t) & = & e^{-\chi_l^2 f(\bar{\omega}, t)}
 \nonumber \\
 & = & e^{-\chi_l^2 A_1}
  \sum_{m,n=0}^\infty \frac{(-1)^{m+n}}{m!\ \! n!}
  \ \! \chi_l^{2(m+n)} A_3^m A_4^n \nonumber \\
 & & \qquad \qquad \qquad \times
  e^{i(m-n)\delta} e^{-(m+n) \eta t/2} e^{i(m-n) \bar{\omega} t} ,
 \label{isfho2}
\ea
where
\be
 f(\bar{\omega}, t) \equiv
  A_1 + A_3 e^{i\delta} e^{-(\eta/2 - i\bar{\omega})t}
  + A_4 e^{-i\delta} e^{-(\eta/2 + i\bar{\omega})t} ,
\ee
with
\ba
 A_1 & = & \frac{\omega_0}{\bar{\omega}}
 \frac{\eta^2 \{2 (\eta/2) \bar{\omega} \sin \delta
  + [\bar{\omega}^2 - (\eta/2)^2] \cos \delta\}}
   {[(\eta/2)^2 + \bar{\omega}^2]^2} ,
 \\
 A_3 & = & \frac{\omega_0}{\bar{\omega}}
  \frac{\eta^2}{2(\eta/2 - i\bar{\omega})^2} ,
 \\
 A_4 & = & \frac{\omega_0}{\bar{\omega}}
  \frac{\eta^2}{2(\eta/2 + i\bar{\omega})^2} ,
\ea
where the coefficients $A_i$ has been put in terms
of $\eta$, $\bar{\omega}$ and $\delta$.
From equation~(\ref{isfho2}), it is now straightforward to derive an
expression for the dynamic scattering factor, which results
\ba
 S(\Delta K, \omega) & = & \frac{e^{-\chi_l^2 A_1}}{\pi}
  \sum_{m,n=0}^\infty \frac{(-1)^{m+n}}{m!\ \! n!}
  \ \! \chi_l^{2(m+n)} A_3^m A_4^n e^{i(m-n)\delta} \nonumber \\
 & & \qquad \qquad \times
   \frac{(m+n) \eta/2}{[\omega - (m-n)\bar{\omega}]^2
    + [(m+n) \eta/2]^2} .
 \label{dsfho2}
\ea
For a harmonic oscillator, there is no diffusion and, therefore,
equation~(\ref{dsfho2}) is only valid when $m \neq n$. All the
Lorentzian functions contributing to equation~(\ref{dsfho2})
are due to the creation and annihilation events of the T mode.
These Lorentzians are characterized by a width given by
$\Gamma = (m+n) \eta/2$, which increases with $\eta$.
This broadening (proportional to $\eta$) undergone by the dynamic
structure factor is thus contrary to the narrowing effect observed
in the case of a flat surface.
It can be assigned to the confined or bound motion displayed by the
particle ensemble when trapped inside the potential wells.
Hence, in order to detect broadening of the line shapes in surface
diffusion experiments, adparticles must spend some time confined inside
potential wells, since the broadening will be induced by the presence
of temporary vibrational motions.


\subsubsection{General periodic surface potentials.
Temporary trapped adparticles.}
\label{sec2.2.8}

As seen above, the broadening of the diffusion line shapes is provoked
by the temporary trapping.
To demonstrate this assertion here we are going to consider a general
velocity autocorrelation function which keeps the functional form of
equation~(\ref{corrHO2}), but whose parameters do not hold the same
relations as those characterizing a harmonic oscillator \cite{JLvega1}.
That is,
\begin{equation}
 \mathcal{C}(t) = \frac{k_B T}{m} \ \! e^{- \tilde{\eta} \tau}
  \cos (\tilde{\omega}t + \tilde{\delta}) ,
 \label{corrg}
\end{equation}
where the values of the parameters $\tilde{\eta}$, $\tilde{\omega}$ and
$\tilde{\delta}$ are obtained from a fitting to the numerical results
---there is no any relation among them as in the case
of the harmonic oscillator model.
From equation~(\ref{corrg}) one easily reaches the corresponding
expression for the intermediate scattering function,
\ba
 I(\Delta K, t) & = & e^{-\chi_l^2 \tilde{f}(\tilde{\omega},t)}
 \nonumber \\
 & = & e^{-\chi_l^2 \tilde{A}_1 - \chi_l^2 \tilde{A}_2 t}
  \sum_{m,n=0}^\infty \frac{(-1)^{m+n}}{m! \ \! n!} \ \!
  \chi_l^{2(m+n)} \tilde{A}_3^m \tilde{A}_4^n \nonumber \\
 & & \qquad \qquad \qquad \times
  e^{i(m-n)\tilde{\delta}} e^{-(m+n)\tilde{\eta}t}
   e^{i(m-n)\tilde{\omega}t} ,
 \label{isfg}
\ea
which is analogous to equation~(\ref{isfho2}). In equation~(\ref{isfg}),
\be
 \tilde{f}(\tilde{\omega},t) \equiv \tilde{A}_1 + \tilde{A}_2 t
  + \tilde{A}_3 e^{i\tilde{\delta}} e^{-(\tilde{\eta}
   - i\tilde{\omega})t}
  + \tilde{A}_4 e^{-i\tilde{\delta}} e^{-(\tilde{\eta}
   + i\tilde{\omega})t} ,
\ee
and
\ba
 \tilde{A}_1 & = & \frac{\tilde{\eta}^2
  [2 \tilde{\eta} \tilde{\omega} \sin \tilde{\delta}
  + (\tilde{\omega}^2 - \tilde{\eta}^2) \cos \tilde{\delta})}
  {(\tilde{\eta}^2 + \tilde{\omega}^2)^2} ,
 \\
 \tilde{A}_2 & = &
  \frac{\tilde{\eta}^2 (\tilde{\eta} \cos \tilde{\delta}
   - \tilde{\omega} \sin \tilde{\delta})}
  {\tilde{\eta}^2 + \tilde{\omega}^2} ,
 \\
 \tilde{A}_3 & = & \frac{\tilde{\eta}^2}
  {2(\tilde{\eta} - i\tilde{\omega})^2} ,
 \\
 \tilde{A}_4 & = & \frac{\tilde{\eta}^2}
  {2(\tilde{\eta} + i\tilde{\omega})^2} .
\ea
Unlike the case of the harmonic oscillator, notice now that there is a
linear dependence on time in $\tilde{f}$ because of the independence
of $\tilde{\eta}$, $\tilde{\omega}$ and $\tilde{\delta}$.
This leads to an exponentially decaying factor in
equation~(\ref{isfg}), which accounts for the diffusion and that makes
the intermediate scattering function to vanish at asymptotic times.
In this sense, the intermediate scattering function can be considered
as containing both phenomena, diffusion and low vibrational motions.
This effect is better appreciated in the dynamic structure factor,
\ba
 S(\Delta K, \omega) & = & \frac{e^{-\chi_l^2 \tilde{A}_1}}{\pi}
  \sum_{m,n=0}^\infty \frac{(-1)^{m+n}}{m!\ \! n!}
  \ \! \chi_l^{2(m+n)} \tilde{A}_3^m \tilde{A}_4^n e^{i(m-n)\delta}
 \nonumber \\
 & & \qquad \times
   \frac{\chi_l^2 \tilde{A}_2 + (m+n) \tilde{\eta}}
   {[\omega - (m-n)\tilde{\omega}]^2
    + [\chi^2 \tilde{A}_2 + (m+n) \tilde{\eta}]^2} .
 \label{dsfg}
\ea
This general expression clearly shows that both motions (diffusion and
oscillatory) cannot be separated at all.
The Q--peak is formed by contributions where $m=n$, for which each
partial FWHM is given by
\be
 \Gamma_Q = \chi_l^2 \tilde{A}_2 + 2 \, m \, \tilde{\eta}.
\ee
Analogously, the T--peaks come from the sums with $n \neq m$ and each
partial FWHM is given by
\be
 \Gamma_T = \chi_l^2 \tilde{A}_2 + (m + n) \, \tilde{\eta}.
\ee
If the Gaussian approximation is good enough,
the value of $\tilde{\eta}$ will not be too different from
the nominal value of $\eta$ and, therefore, both peaks
will display broadening as $\eta$ increases.
This is a very remarkable result since a relatively simple model,
as the one described here, can explain the corresponding experimental
broadenings observed with coverage. Thus, broadening arises from the
temporary confinement of the adparticles inside potential wells along
their motion on the surface \cite{prl-I}. The problem of the
experimental deconvolution has been already discussed elsewhere
\cite{jcp-elliot}. Here we would like only to mention that using this simple
model, such deconvolutions would be more appropriate in order to
extract useful information about diffusion constants and jump mechanisms.
Finally, as mentioned before, the motional narrowing effect will
govern the gradual change of the whole line shape as a function of
the friction or, equivalently, the coverage, the parallel momentum
transfer and the jump mechanism.


\section{Results}
\label{sec3}

For a better understanding of the concepts introduced in
section~\ref{sec2}, here we present results for two different
types of two-dimensional surfaces: flat and corrugated.
Since much research has been developed for Na atoms adsorbed on a
Cu(001) surface, we will carry out our numerical simulations taking
into account the periodic, separable corrugated potential found in
the literature for this system (see, for example,
reference~\cite{toennies2}).


\subsection{Computational details}
\label{sec3.1}

As in reference~\cite{toennies2}, we have considered two values of
the coverage in our calculations, $\theta = 0.028$ and 0.18, where
$\theta_{Na} = 1$ corresponds to one Na atom per Cu(001) surface atom
---or, equivalently, $\sigma = 1.53 \times 10^{19}$~atom/cm$^2$
\cite{toennies3}. All the friction coefficients are given in atomic
units.
Taking into account the values $a = 2.557$~\AA\ and $\rho = 2$~\AA\
for the unit cell length and the Na atomic radius, and using
equation~(\ref{theta}), the values of $\lambda$ associated with the
coverages used here are $\lambda = 3.34\times10^{-6}$ a.u.\ for
$\theta = 0.028$ and $\lambda = 2.15\times10^{-5}$ a.u.\ for
$\theta = 0.18$ at $T = 200$~K.
For the surface friction we have considered the value also given in the
literature \cite{toennies2}, $\gamma = 0.1 \ \! \omega_0 = 2.2049\times
10^{-5}$ a.u., where $\omega_0$ is the harmonic frequency associated
with the periodic surface potential (see section~\ref{sec3.3}).
With these two frictions ($\lambda$ and $\gamma$) the resulting total
friction is $\eta \approx 2.53\times10^{-5}$ a.u.\ for $\theta = 0.028$
and $\eta \approx 4.35\times10^{-5}$ a.u.\ for $\theta = 0.18$.
Regarding the collision relaxation rate, we have assumed $\lambda' =
10^{-3}$ a.u., which allows us to use the standard Langevin equation
with a white shot noise.
As far as we know, no information about that parameter is found in
the literature and therefore the interacting single adsorbate
approximation could also give us an estimation of the duration of the
adsorbate-adsorbate collision.
Finally, the Verlet algorithm \cite{allen} is used to solve the
corresponding Langevin equation.


\subsection{Flat surface model. The low corrugation regime}
\label{sec3.2}

First we analyze the case of adsorbate diffusion on a flat surface,
i.e., the case of a two-dimensional gas \cite{prb-I,toeprl}.
This example is representative of low corrugated surfaces, where the
role of the activation barrier is negligible.

\begin{figure}
 \begin{center}
 \epsfxsize=4in {\epsfbox{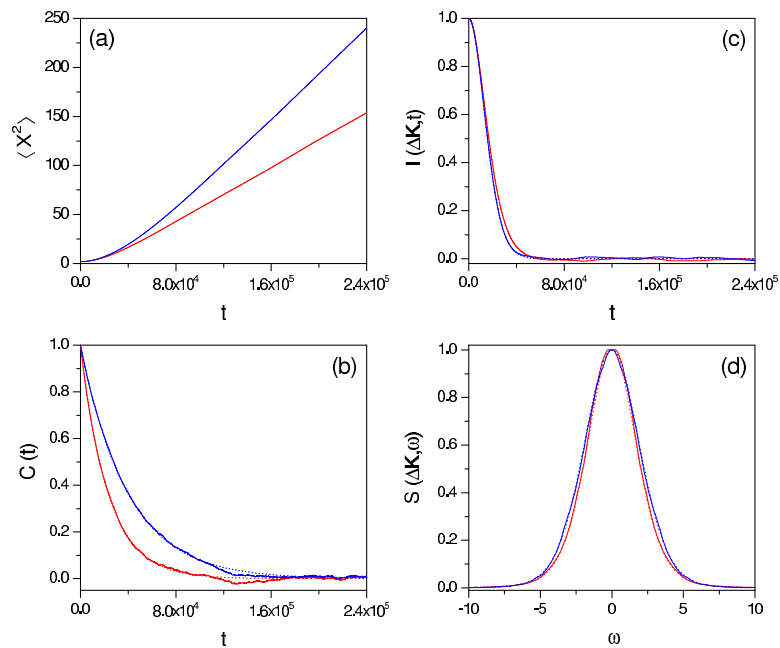}}
 \caption{\label{fig4}
  Dynamical magnitudes for two different values
  of the coverage, $\theta=0.028$ (blue) and $\theta=0.18$ (red):
  (a) $\langle x^2(t) \rangle$, (b) $\mathcal{C}(t)$,
  (c) $I(\Delta {\bf K},t)$ and (d) $S(\omega)$, with
  $\Delta K = 1.23$~\AA$^{-1}$.
  Dotted lines are the numerical fittings to the corresponding
  analytical formulas given in section~\ref{sec2}.}
 \end{center}
\end{figure}

As seen in section~\ref{sec2.2.4}, when $V = 0$, the dynamical
magnitude $\langle x^2(t) \rangle$ displays two different dynamical
regimes: ballistic ($t \ll 1/\eta$) and diffusive ($t \gg 1/\eta$).
This can be seen in figure~\ref{fig4}(a), where $\langle x^2(t)\rangle$
is effectively proportional to $t$ in the long time regime and depends
on $t^2$ at short times (of the order of $1/\eta$).
In the linear regime, the slope of $\langle x^2(t) \rangle$ increases
as $\eta$ decreases, i.e., in accordance with Einstein's law (see
equation~(\ref{avvalues5})), the diffusion decreases with the friction.
By fitting the linear part of the graphs (dotted lines) to
equation~(\ref{avvalues5}), we get $D = 6.045\times 10^{-4}$ a.u.\
for $\theta = 0.028$ and  $D = 3.510\times 10^{-4}$ a.u.\ for
$\theta = 0.18$.
Note that these diffusion coefficients are not only related to the
friction due to the surface (i.e., the adsorbate-substrate coupling),
but also to the collisions among adsorbates.
Indeed, we can see that for a given surface friction, the diffusion
is {\it inhibited} by such collisions ---it {\it decreases} with the
coverage.
This is a very remarkable result since in real experiments the surface
friction is fixed and diffusion can then be studied taking only into
account the coverage of the surface.
On the other hand, from these values for $D$ we obtain, respectively,
$\eta = 2.50\times 10^{-5}$ a.u.\ and $\eta = 4.35\times 10^{-5}$
a.u., which are in a good agreement with those used in our calculations
(see section~\ref{sec3.1}).
This agreement between the simulation and the analytical model is
particularly important because it validates our standard Langevin
description.

The effect of the collisions also produces  an important effect on
the velocity autocorrelation function, $\mathcal{C}(t)$, plotted in
Fig.~\ref{fig4}(b).
As seen, as $\theta$ increases this function decays faster.
In particular, from the fitting of these results to
equation~(\ref{corrGM2}) (normalized to unity) we have obtained
$\eta \approx 2.52\times10^{-5}$ a.u.\ for $\theta = 0.028$ and
$\eta \approx 4.32\times10^{-5}$ a.u.\ for $\theta = 0.18$, which
again show a good agreement with the actual values employed in our
simulations.
This fast decay also indicates that higher order correlations (e.g.,
correlations at three or four times) will decay much faster.
Their effect will be then negligible on the intermediate scattering
function, thus validating the use of the Gaussian approximation,
since such higher order correlations will not be relevant when
passing from equation~(\ref{eq:IntSF}) to (\ref{eq:IntSF2}).
The validity of the Gaussian approximation can be seen more explicitly
when comparing the intermediate scattering functions obtained from
the calculations with that fitted using equation~(\ref{eq:IntSGM}).
These results are plotted in figure~\ref{fig4}(c) for
$\Delta K = 1.23$~\AA$^{-1}$, where the fitting has been carried out
with the values of $\eta$ used in the simulation and considering
$\chi$ as the fitting parameter.
We observe that not only the correspondence between the simulations
and the analytical formulas are excellent from the figure, but also
from the fitted values of $\chi$: $\chi_{fit} = 3.16$ vs
$\chi_{sim} = 3.16$ for $\theta = 0.028$, and $\chi_{fit} = 1.81$
vs $\chi_{sim} = 1.83$ for $\theta = 0.18$.
Notice that in agreement with equation~(\ref{eq:IntSGM}), $I(t)$
displays an initial Gaussian falloff at short times, while for longer
times its decay is exponential.
Moreover, as also expected from equation~(\ref{eq:IntSGM}), and
equation~(\ref{eq:chi2}), as the coverage increases a slower decay
is observed.

Finally, in figure~\ref{fig4}(d) we have plotted the dynamic structure
factor after time Fourier transforming the intermediate scattering
function obtained from both the numerical calculations and their
corresponding analytical fittings. Note that, a narrowing of the
Q peak is predicted with the coverage \cite{prl-I}.
Regarding the shape of this peak, it can be shown that it is a
mixture of Gaussian and Lorentzian functions.
The Gaussian behaviour is ruled by the short time limit of the
intermediate scattering function, while the Lorentzian behaviour
arises from the long time exponential decay.
The shape of the Q peak will depend
on which of the two extreme regimes is dominant (motional narrowing
effect) \cite{JLvega1}.
This analysis should be carried out when experimental results are
deconvoluted \cite{prb-I} since it is very common to see fittings of
the Q peak to a pure Lorentzian function.


\subsection{Periodic, corrugated surface model}
\label{sec3.3}

Now we are going to study the case of a periodic surface model
described by the potential
\be
 V (x,y) = V_0 \left[2-\cos(2\pi x/a) - \cos(2\pi y/a) \right] ,
 \label{eq:Pot0}
\ee
with $a=2.557$~\AA\ and $V_0=37.32$~meV.
In this case, the surface corrugation is relatively strong and
cannot be neglected, thus inducing important effects in the adsorbate
dynamics.
These effects can be well understood by using the third model
described in section~\ref{sec2.2.8}.

\begin{figure}
 \begin{center}
 \epsfxsize=4.25in {\epsfbox{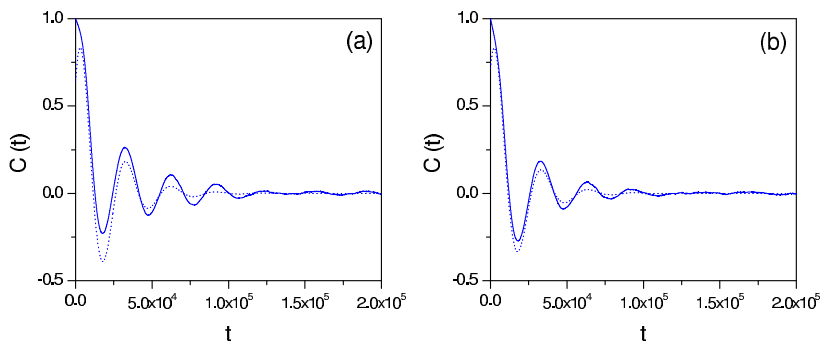}}
 \caption{\label{fig5}
  Velocity autocorrelation function, $\mathcal{C}(t)$,
  for two different values of the coverage: (a) $\theta=0.028$ and (b)
  $\theta=0.18$.
  Solid lines denote the numerical results obtained from the
  simulations and dotted lines the fittings to equation~(\ref{corrg}).}
 \end{center}
\end{figure}

After figures~\ref{fig5}, the diffusive and trapped regimes that
appear under the presence of the surface corrugation give rise to
the oscillating behaviour of the velocity autocorrelation function.
The adparticles trapped inside the potential wells give rise to the
oscillations, while the exponential damping arises as a consequence
of the loss of correlation with time of both the particles moving
outside the wells and those trapped inside.
This behaviour thus approaches that described by equation~(\ref{corrg}),
as seen from the fittings (dotted line) with parameters $\tilde{\eta}
= 5.13\times 10^{-5}$ a.u., $\tilde{\omega} = 2.14\times 10^{-4}$ a.u.\
and $\tilde{\delta} = -0.874$ for $\theta = 0.028$ (see
figure~\ref{fig5}(a)), and $\tilde{\eta} = 5.96\times 10^{-5}$ a.u.,
$\tilde{\omega} = 2.05\times 10^{-4}$ a.u.\ and $\tilde{\delta} =
-0.776$ for $\theta = 0.18$ (see figure~\ref{fig5}(b)).
Note that, since we have contributions from running and trapped or
bound trajectories, here the fitting is not so good as that seen in
section~\ref{sec3.2} for $V = 0$.
Recently, it has been shown \cite{prl-I} that if the velocity
autocorrelation function is expressed as a linear combination of the
corresponding functions for free particle and anharmonic oscillator,
the fitting is highly improved. Obviously, the Gaussian approximation
is no longer valid due to the presence of the corrugated potential.
Nonetheless, it is important to stress that the model given by
equation~(\ref{corrg}) is still valid to interpret the results
obtained; not only the fitted parameters are of the order of those
employed in the calculations, but they also follow the trend that one
might expect from the exact ones.
That is, as $\tilde{\eta}$ increases the exponential damping in
$\mathcal{C}(t)$ is stronger and a slight shift of the oscillations
is observed.

\begin{figure}
\begin{center}
 \epsfxsize=4in {\epsfbox{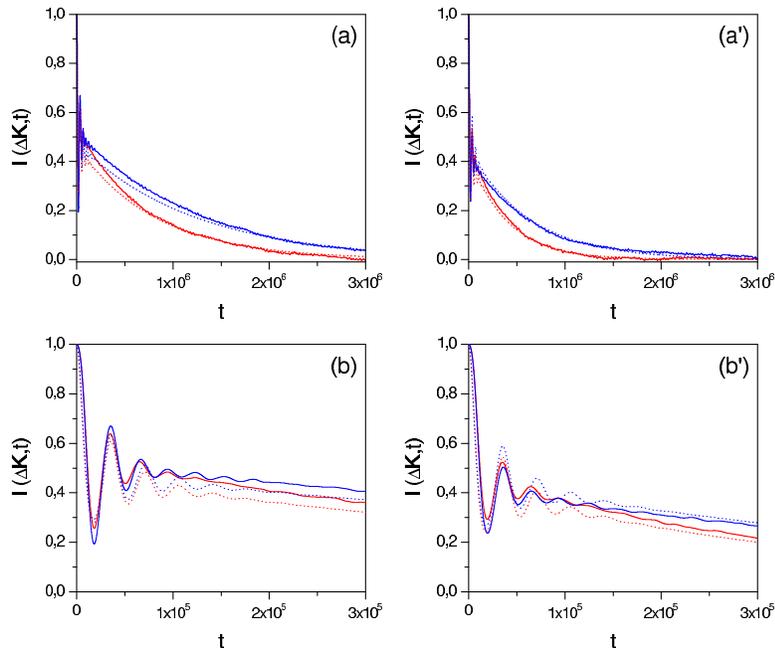}}
 \caption{\label{fig6}
  Intermediate scattering function for
  $\Delta K = 1.23$~\AA$^{-1}$ and two different values of the
  coverage: $\theta=0.028$ in panels (a) and (a') along two
  directions, [110] and [100] and $\theta=0.18$ in
  panels (b) and (b') along the same two directions.
  In the bottom panels observe details of the short time regime.
  Solid lines denote the numerical results obtained from the
  simulations and dotted lines the fittings to equation~(\ref{isfg}).}
 \end{center}
\end{figure}

As seen in figure~\ref{fig6}, the effects of the trapping or bound
motion are also apparent in the intermediate scattering function,
which is plotted for $\Delta K = 1.23$~\AA$^{-1}$, along two different
directions and two coverages: panels (a) and (a') for $\theta = 0.028$
along the [110] and [100] directions, respectively; and panels (b)
and (b') for $\theta = 0.18$ and along the same two directions.
Note first that again here we can distinguish two decay regimes (which
are present along both directions). The long time regime in the upper
panels is characterized by an exponential
damping, as in the case for $V = 0$, which is typical of diffusion and
will give rise to a predominance of the Lorentzian-like behaviour in
the dynamic structure factor (see below).
On the other hand, the short time regime in the bottom panels clearly
shows the damped oscillating behaviour of $\mathcal{C}(t)$.
These oscillations at short times will give rise to the T-mode
peaks associated with the low frequency vibrational motions induced
by the potential wells in the trapped trajectories.
Moreover, the oscillating behaviour is more patent with the coverage
giving rise to an {\it inversion} in the behaviour of the long
time tail of $I(t)$. Unlike the intermediate scattering function
associated with a flat surface dynamics, this faster decay with
coverage will lead to a broadening of the Q peak in the dynamic
structure factor.
Moreover, also notice that the timescales ruling the decay are about
two orders of magnitude larger than in the case with $V = 0$.
This is due to the presence of the potential, which allows to keep
the phase correlation in equation~(\ref{eq:IntSF}) for longer times
due to the trapping inside the potential wells \cite{prl-I}.

\begin{figure}
\begin{center}
 \epsfxsize=4in {\epsfbox{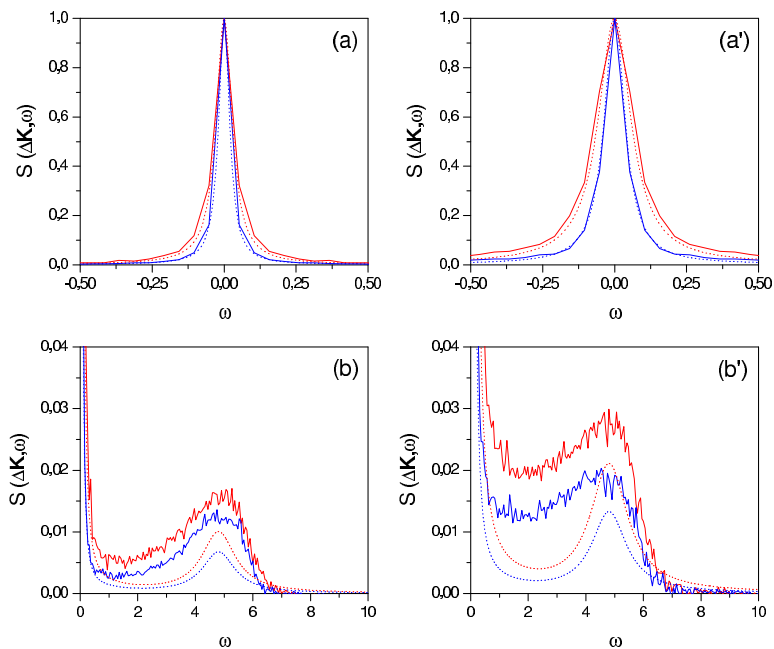}}
 \caption{\label{fig7}
  Dynamic structure factor for
  $\Delta K = 1.23$~\AA$^{-1}$ and two different values of the
  coverage: $\theta=0.028$ in panels (a) and (a') along two
  directions, [110] and [100] and $\theta=0.18$ in panels (b)
  and (b') along the same two directions.
  In the bottom panels observe details of the peaks corresponding to
  the T mode.
  Solid lines denote the numerical results obtained from the
  simulations and dotted lines the time Fourier transform of the
  fittings to equation~(\ref{isfg}).}
 \end{center}
\end{figure}

Regarding the dynamic structure factor, displayed in
figure~\ref{fig7} (as before, the four panels correspond to the same
coverages and directions as in figure~\ref{fig6}),
the presence of the potential has three
important differences with respect to the flat surface case.
First, the trapping trajectories due to the surface corrugation
give rise to the appearance of two symmetric peaks around the
Q peak, the T-mode peaks which are located at
$-5$~meV and 5~meV, approximately (in the lower panels of
figure~\ref{fig7} only the right peak is showed); these peaks broadens
with coverage.
Second, the faster decay of $I(t)$ as the coverage increases translates
into a broadening of the Q peak, which is in accordance to what one
can observe experimentally \cite{toennies2}.
Third, the Q peak is narrower than the corresponding in the flat
surface case, around two orders of magnitude.
This narrowing is due to the {\it inhibition} of the diffusive motion
induced by the corrugation of the potential (a large number of
particles get trapped inside the wells).
In any case, the diffusion coefficient is governed by the Einstein
relation and therefore decreases with the coverage.

Finally, it is worth stressing that there is a strong correlation
between diffusion and low frequency vibrational motions \cite{JLvega1}.
This fact is easily appreciated in the two bottom panels of
figure~\ref{fig7}, where a strong overlapping of the two peaks is
noticeable.
In the Gaussian approximation, this overlapping can be understood
from equation~(\ref{dsfg}).
In general, in order to extract the width of the Q and T peaks,
experimentalists usually follows a deconvolution procedure where an
effective Lorentzian function is assumed for each peak independently.
After the study presented here (as well as in
references~\cite{JLvega1,jcp-elliot} within the single adsorbate
approximation), this deconvolution procedure could lead to
inconsistencies when extracting information because both peaks are
strongly overlapped.
Indeed, it can be shown \cite{jcp-I} that depending on the parallel
momentum transfer $\Delta {\bf K}$ considered, the overlapping of both
peaks may become so strong that the Lorentzian fitting procedure would
not be valid anymore.
In the light of this study we propose a more general fitting procedure
based on equation~(\ref{dsfg}).


\section{Conclusions}
\label{sec4}

In this work, we have presented a full  stochastic description
of surface diffusion and low frequency vibrational motion for
adsorbate/substrate systems with increasing coverages.
A Langevin equation is used to describe the adparticle dynamics,
where a Gaussian white noise  simulates the effects of
adsorbate-substrate friction and a white shot noise  is used
for the adsorbate-adsorbate interactions within what we call the
interacting single adsorbate approximation.
As shown, this theoretical stochastic framework not only provides a
simple and complementary view of the processes taking place on a surface,
but also explains the experimental observations, i.e., the line shape
broadening of the Q peak ruling the diffusion process as a function
of the coverage.
The idea of replacing the dipole-dipole interaction by a white shot
noise turns out to be crucial because for long time processes, as the
diffusion one, a high number of collisions occurs and the statistical
limit seems to wipe out any trace of the true interaction potential.

The main points of this work can be emphasized as follows.
First, the simplicity of the stochastic approach developed here is
in sharp contrast to the Langevin molecular dynamics simulations
commonly used to study adsorbate diffusion.
This treatment allows to derive analytical formulas that render some
light into the physics involved in surface diffusion, something that
is not possible by means of massive molecular dynamics calculations
which provide numbers that cannot be easily handled by using an
analytical data processing methodology.
Second, by using this numerical and analytical treatment, we have been
able to show that the main reason for the broadening of the Q peak is
the trapping induced by the surface corrugation.
Third, this analysis constitutes a robust and simple methodology that
can be of great help for the experimentalists in the interpretation
of their results.
And fourth, the results obtained here can be of relevance in problems
where friction has to be varied gradually.
Note that the surface friction is a fixed parameter, since it depends
only on the type of surface.
However, relating the coverage with a friction ($\lambda$) seems to be
an appropriate manner of studying such variations.

Finally, this class of studies can obviously not replace further
investigation at microscopic level and calculations from first
principles of adsorbate dynamics.
For example, the Lau-Kohn long range interaction
via intrinsic surface states \cite{giorgio} has been proposed to use a
coverage dependent surface electronic density modifying the corrugated
potential. It has also been used to explain the remarkable increasing
of the T-mode frequency with coverage and where the dipole-dipole
interaction is not able to reproduce this behaviour. Our simple
stochastic model provides a complementary view of the diffusion and
low-frequency vibrational motion features observed as peaks around
or near zero energy transfers (or long time regime) in the scattering
law.


\ack

This work was supported in part by DGCYT (Spain) under the project
with reference number FIS2004-02461.
R.M.-C.\ would like to thank the University of Bochum for support
from the Deutsche Forschungsgemeinschaft, SFB 558, for a predoctoral
contract.
J.L.V.\ and A.S.\ Sanz would like to thank the Spanish Ministry of
Education and Science for a predoctoral grant and a ``Juan de la
Cierva'' Contract, respectively.



\Bibliography{99}

\bibitem{Gomer}
 Gomer R 1990 {\it Rep. Prog. Phys.} {\bf 53} 917

\bibitem{Ehrlich}
 Ehrlich G 1994 {\it Surf. Sci.} {\bf 300} 628

\bibitem{toennies4}
 Hofmann F and Toennies J P 1996 {\it Chem. Rev.} {\bf 78} 3900

\bibitem{salva}
  Miret-Art\'es S and Pollak E 2005
 {\it J. Phys.: Condens. Matter} {\bf 17} S4133

\bibitem{jardine1}
 Jardine A P, Dworski S, Fouquet P, Alexandrowicz G, Riley D J,
 Lee G Y H, Ellis J and Allison W 2004 {\it Science} {\bf 104} 1790;
 \newline
 Fouquet P, Jardine A P, Dworski S, Alexandrowicz G, Allison W
 and Ellis J 2005 {\it Rev. Sci. Inst.} {\bf 76} 053109

\bibitem{toennies1}
 Graham A P, Hofmann F and Toennies J P 1996
 {\it J. Chem. Phys.} {\bf 104} 5311

\bibitem{toennies2}
 Graham A P, Hofmann F, Toennies J P, Chen L Y and Ying S C 1997
 {\it Phys. Rev. Lett.} {\bf 78} 3900; \newline
 Graham A P, Hofmann F, Toennies J P, Chen L Y and Ying S C 1997
 {\it Phys. Rev. B} {\bf 56} 10567

\bibitem{toennies3}
 Ellis J, Graham A P, Hofmann F and Toennies J P 2001
 {\it Phys. Rev. B} {\bf 63} 195408

\bibitem{vanhove}
 Van Hove L, 1954 {\it Phys. Rev.} {\bf 95} 249

\bibitem{vineyard}
 Vineyard G H, 1958 {\it Phys. Rev.} {\bf 110} 999


\bibitem{JLvega0}
 Vega J L, Guantes R and Miret-Art\'es S 2002
 {\it J. Phys.: Condens. Matter} {\bf 14} 6193

\bibitem{JLvega1}
 Vega J L, Guantes R and Miret-Art\'es S 2004
 {\it J. Phys.: Condens. Matter} {\bf 16} S2879

\bibitem{eli}
 Guantes R, Vega J L, Miret-Art\'es S and Pollak E 2003
 {\it J. Chem. Phys.} {\bf 119} 2780; \newline
 Guantes R, Vega J L, Miret-Art\'es S and Pollak E 2004
 {\it J. Chem. Phys.} {\bf 120} 10768

\bibitem{sancho}
 Sancho J M, Lacasta A M, Lindenberg K, Sokolov I M and Romero A H
 2004 {\it Phys. Rev. Lett.} {\bf 92} 250601

\bibitem{prl-I}
 Mart\'{\i}nez-Casado R, Vega J L, Sanz A S and Miret-Art\'es S
 2007 {\it Phys. Rev. Lett.} {\bf 98} 216102
 ({\it Preprint} cond-mat/0702219)

\bibitem{pre-I}
 Mart\'{\i}nez-Casado R, Vega J L, Sanz A S and Miret-Art\'es S
 2007 {\it Phys. Rev. E} {\bf 75} 051128
 ({\it Preprint} cond-mat/0608723)

\bibitem{vvleck-weiss}
 van Vleck J H and Weisskopf V F 1945 {\it Rev. Mod. Phys.}
  {\bf 17} 227

\bibitem{mcquarrie}
 McQuarrie D A 1976 {\it Statistical Mechanics}
 (New York: Harper and Row)

\bibitem{gardiner}
 Gardiner C W 1983 {\it Handbook of Stochastic Methods}
 (Berlin: Springer-Verlag)

\bibitem{hanggi1}
 Czernik T, Kula J, Luczka J and H\"anggi P 1997
 {\it Phys. Rev. E} {\bf 55} 4057; \newline
 Luczka J, Czernik T and H\"anggi P 1997
 {\it Phys. Rev. E} {\bf 56} 3968

\bibitem{iturbe}
 Laio F, Porporato A, Ridolfi L and Rodriguez-Iturbe I 2001
 {\it Phys. Rev. E} {\bf 63} 036105

\bibitem{tommei}
 Ferrando R, Mazroui M, Spadacini R and Tommei G E 2005
 {\it New J. Phys.} {\bf 7} 19

\bibitem{hanggi2}
 H\"anggi P and Jung P 1995 {\it Adv. Chem. Phys.} {\bf 89} 239

\bibitem{Hansen}
 Hansen J P and McDonald I R 1986 {\it Theory of simple liquids}
 (London: Academic Press)

\bibitem{topping}
 Topping J 1927 {\it Proc. R. Soc. London}, Ser. A {\bf A114} 67

\bibitem{kubo}
 Kubo R 1966 {\it Rep. Prog. Phys.} {\bf 29} 255

\bibitem{shot-noise}
 Heer C V 1972 {\it Statistical Mechanics, Kinetic Theory, and
 Stochastic Processes} (New York: Academic Press)

\bibitem{schottky}
 Schottky W 1918 {\it Ann. Phys. (Leipzig)} {\bf 57} 541

\bibitem{rice}
 Rice S O 1944 {\it Bell Syst. Tech. J.} {\bf 23} 282; \newline
 Rice S O 1945 {\it Bell Syst. Tech. J.} {\bf 24} 46

\bibitem{prb-I}
 Mart\'{\i}nez-Casado R, Vega J L, Sanz A S and Miret-Art\'es S
 2007 {\it J. Phys.: Cond. Matter} {\bf 19} 176006
 ({\it Preprint} cond-mat/0608724)

\bibitem{jcp-elliot}
 Mart\'{\i}nez-Casado R, Vega J L, Sanz A S and Miret-Art\'es S
 2007 {\it J. Chem. Phys.} {\bf 126} 194711
 ({\it Preprint} cond-mat/0609249)

\bibitem{toeprl}
 Ellis J, Graham A P and Toennies J P 1999
 {\it Phys. Rev. Lett.} {\bf 82} 5072

\bibitem{elliott}
 Chudley C T and Elliott R J 1961 {\it Proc. Phys. Soc.} {\bf 57} 353

\bibitem{risken}
 Risken H 1984 {\it The Fokker-Planck Equation}
 (Berlin: Springer-Verlag)

\bibitem{allen}
 Allen M P and Tildesley D J 1990 {\it Computer Simulation of Liquids}
 (Oxford: Clarendon Press)

\bibitem{jcp-I}
 Mart\'{\i}nez-Casado R, Vega J L, Sanz A S and Miret-Art\'es S
 (to be submitted)

\bibitem{giorgio}
 Graham A P, Toennies J P and Benedel G 2004 {\it Surf. Sci.} L143
\endbib
\end{document}